# A coupled approach to model the effect of wear on the dynamics of the shrouded bladed disk

Lakshminarayana Reddy Tamatam[1]*, Daniele Botto[1] *and* Stefano Zucca[1]

## Abstract

This paper deals with modelling the effect of wear on the dynamics of the shrouded bladed disk with frictional contacts at the shrouds and the contact interface evolution. Prediction of fretting wear commonly occurring at the contacts of turbomachinery components, and its impact on the dynamics is increasingly researched due to the components subjected to their structural limits for performance and operating at high loading conditions. Over a lifetime, the fretting wear at these contacts could alter the global dynamic response of these bladed disks from the designed operating point and could lead to high vibration amplitudes. This study implements a coupled static/dynamic harmonic balance method (HBM) with wear energy approach and an adaptive wear logic to study the impact on the steady-state nonlinear dynamic response. Firstly, the methodology is applied to a cantilever beam with a contact patch and then to a shrouded bladed disk with shroud contacts using cyclic symmetry boundary conditions. The novelty of the paper is to show the effect of wear on the dynamics with the changing contact pre-load using a coupled approach. A wear acceleration parameter $\mathbf{v}_{w,max}$ is defined, and the impact of the choice of this parameter is discussed in detail on the computation time and the accuracy of results. The test cases demonstrate the impact of wear on the dynamic nonlinear response curves and the contact interface evolution for various scenarios.

Keywords: non-linear structural dynamics, contact evolution, contact pre-load, fretting wear, harmonic balance method

## 1. Introduction

Mechanical assemblies made of two or more components have contact interfaces between them. Many a time relative sliding occurs during operation, sometimes intentional for design and functional purposes and at times unintentional. In turbomachinery components such as bladed disks, the relative sliding at blade roots, shroud contacts and under platform dampers are carefully designed to provide friction damping and to reduce the vibration amplitudes. These components have very high modal density, and resonances are unavoidable within the operating range. The contacts are precisely designed to undergo small relative motion and provide friction damping and reduce the peak vibration amplitudes. Consequently, this high-frequency small-amplitude motion leads to fretting wear during operation. This wear affects the global dynamics over a period of time and causes an undesirable shift in resonant frequency and amplitude. The turbomachinery components are designed to perform at its structural limits to obtain the maximum efficiency. Hence, they experience very high static and dynamic loads due to centrifugal forces, thermal strains and fluctuating gas forces, leading to high cycle fatigue (HCF). HCF is one of the common failure modes of the bladed disks. This leads to degraded performance, sometimes even catastrophic failure.

Wear is a multi-physics and multi-scale phenomenon. The micro-slip or gross tangential slip occurring at the contact interfaces leads to contact hysteresis and energy dissipation and cause wear [1], [2]. There is ample literature available on characterizing and predicting fretting wear occurring at these contacts for various contact conditions ([3]–[9]). This microscopic level surface changes can alter the dynamic system response in a significant way even for a relatively simple system, as shown experimentally in Fantetti et al. [10]. The friction contacts introduce nonlinearity to the system. To perform dynamic analysis considering the nonlinearity, one needs robust iterative solvers and accurate contact models to solve the differential equations of motion with friction nonlinearity. The two predominant methods are time integration methods ([11], [12]) and frequency-based methods ([11], [13]–[19]). Harmonic balance method (HBM) is the most widely used frequency-based method to solve the non-linear differential equations due to its suitability for large systems, computational efficiency and directly obtaining steady-state response ([11], [13]–[18]).

---

[1] *Department of Mechanical and Aerospace Engineering,*
*Politecnico di Torino, Corso Duca degli Abruzzi 24, 10129 Torino, Italy*
* *e-mail:* <u>lakshminarayana.tamatam@polito.it</u>

Today's need is embedding the accurate wear prediction tool into these non-linear dynamic solvers to study the impact of wear and contact interface evolution. There is little research available on this joint topic. Brake's book [20] provides an up-to-date comprehensive overview of the developments of mechanics of jointed structures so far. Some of the other notable researches are [10], [21]–[28]. Sfantos et al. ([3], [4]) have proposed the Boundary Element Method (BEM) techniques for wear simulation and optimization methods using incremental sliding BEM. Salles et al. ([24]–[26]) proposed DLFT and HBM techniques to compute the effect of wear on the dynamics. They also further proposed a multi-scale method to predict the effect of fretting wear as the amount of wear in each fretting cycle is too small relative to the change in dynamics. A multi-scale approach would work on two different time scales for wear and the dynamics. Armand et al. ([21], [22]) extended the previous work. They proposed a multi-scale approach in time and space to predict the effect of fretting wear on the dynamic response for bladed disk under platform dampers application. Later, they also modelled the contact considering surface roughness and showed the change in the dynamic response of the system [23]. Albeit all these studies are only numerical so far.

To the author's knowledge, the only research that studied the effect of wear on the evolution of dynamics of friction contact experimentally and characterized by a constitutive numerical model was by Fantetti et al. [10]. Fantetti et al. showed the impact of wear on the dynamics of structures experimentally and then validated using a constitutive numerical model - modified Bouc-Wen model. They studied the evolution of fretting wear and the dynamics during the running-in up to steady-state. They captured the evolution of fretting wear for a very short number of cycles up to a few millions of cycles. The study clearly showed the impact of the microscopic wear generated at the contact has on the resonant frequencies and damping and the corresponding change in dynamics in the running-in and steady-state conditions. The authors consider this research as the gateway to understand the effect of wear on the dynamics and for future predictive modelling. Overall, there is a need for a reliable and high-fidelity predictive model to include wear and to predict the impact of wear on the global dynamics of the systems.

The current research is a numerical work and aims at developing and implementing a method to predict the effect of wear on the dynamics of structures with contact interfaces and also the evolution of contact interface using a proven static/dynamic coupled approach with harmonic balance method (HBM) [27]. Concerning the previous works existing in the literature ([29], [30]), the current approach allows the automatic update of the static pre-load distribution over the contact area during the non-linear dynamic analysis, without any need for a separate static analysis. The coupled approach is more relevant in shrouded blades, where the static pre-load at the contacts is strongly affected by the worn-out material at the shrouds, hence the vibration amplitudes. This is different from what happens at the under-platform dampers and blade root joints, where wear mostly affects the static load distribution. The wear is computed using wear energy approach [31], and an accelerated adaptive wear logic is defined. Two test cases are chosen to demonstrate the method – a cantilever beam with a contact patch and a shrouded bladed disk with shroud contacts and cyclic symmetry boundary conditions.

The novelty of this paper is two-fold:

1) Modelling the effect of wear on the dynamic forced response and the contact interface evolution with 'changing contact pre-load'.

2) The impact of the choice of user-defined wear acceleration parameter $\mathbf{v}_{w,max}$ that is described in the paper.

## 2. Methodology

The contact interface introduces nonlinearity to the system of equations. Two ways to solve the non-linear differential equation is using time domain and frequency domain methods. A time-domain method such as Direct Time Integration (DTI) provides the transient as well as steady-state response, but is computationally demanding and is not a feasible solution for practical scenarios and systems with large degrees of freedom (DOF). The state-of-the-art frequency-domain method to compute the non-linear response is the Harmonic Balance Method (HBM). This method assumes a periodic response under periodic excitation. Krack and Gross's book [32] provides a detailed description of the method and examples of non-linear vibration problems. This method offers speedy solution times assuming steady-state response and much less computationally demanding relative to the time domain methods. In our case, this is better as we are interested only in the steady-state response of the system subjected to a periodic excitation.

The current study uses a more recent formulation of the HBM, which allows performing a coupled static and dynamic analysis of the system. This coupled method is proven to provide accurate results in comparison to the classical uncoupled approach [29]. It is comparable to the accuracy of direct time integration results with a sufficient number of harmonics.

### 2.1 Governing equations of a generic system

The governing equation of motion of a generic system with contact interfaces is written as:

$$\mathbf{M}\ddot{\mathbf{Q}}(t) + \mathbf{C}\dot{\mathbf{Q}}(t) + \mathbf{K}\mathbf{Q}(t) = \mathbf{F}(t) + \mathbf{F}_c(\mathbf{Q}, \dot{\mathbf{Q}}, t) \qquad (1)$$

where $\mathbf{M} \in \mathbb{R}^{N \times N}$ is the mass matrix; $\mathbf{C} \in \mathbb{R}^{N \times N}$ is the viscous damping matrix; $\mathbf{K} \in \mathbb{R}^{N \times N}$ is the stiffness matrices; $\mathbf{Q}(t) \in \mathbb{R}^{N \times 1}$ is a nodal displacement vector; $\dot{\mathbf{Q}}$ is the time derivative of $\mathbf{Q}$; $\mathbf{F}(t) \in \mathbb{R}^{N \times 1}$ is the excitation force vector and $\mathbf{F}_c(\mathbf{Q}, \dot{\mathbf{Q}}, t) \in \mathbb{R}^{N \times 1}$ is the nonlinear contact force interaction vector. The components of the displacement vector $\mathbf{Q}$ is given by:

$$\mathbf{Q} = [\mathbf{q}_c^T \ \mathbf{q}_a^T \ \mathbf{q}_o^T]^T \qquad (2)$$

where $\mathbf{q}_c$, $\mathbf{q}_a$ and $\mathbf{q}_0$ are the displacements of contact node DOFs, accessory node DOFs, and other node DOFs of the system respectively. $\mathbf{q}_a$ node DOFs include input and output node DOFs.

Since the contact DOF $\mathbf{q}_c$ and the accessory DOF $\mathbf{q}_a$ are usually much smaller in comparison to the other DOF $\mathbf{q}_o$, a reduction of the size of the nonlinear model is typically performed before computing the nonlinear solution. In this paper, the classical Craig-Bampton reduction method [33] is applied, retaining as physical coordinates the contact DOFs $\mathbf{q}_c$ and accessory DOFs $\mathbf{q}_a$ – cumulatively called as master nodes, and adding a certain number of fixed-interface modes $\boldsymbol{\eta}$.

$$\mathbf{q} = [\mathbf{q}_c^T \ \mathbf{q}_a^T \ \boldsymbol{\eta}^T]^T \qquad (3)$$

where $\mathbf{q} \in \mathbb{R}^{n \times 1}$ and $n \ll N$. The equation (1) can be rewritten as:

$$\mathbf{m}\ddot{\mathbf{q}}(t) + \mathbf{c}\dot{\mathbf{q}}(t) + \mathbf{k}\mathbf{q}(t) = \mathbf{f}(t) + \mathbf{f}_c(\mathbf{q}, \dot{\mathbf{q}}, t) \qquad (4)$$

where $\mathbf{m} \in \mathbb{R}^{n \times n}$ is the mass matrix, $\mathbf{c} \in \mathbb{R}^{n \times n}$ is the viscous damping matrix and $\mathbf{k} \in \mathbb{R}^{n \times n}$ is the stiffness matrix, $\mathbf{q}(t) \in \mathbb{R}^{n \times 1}$ is a nodal displacement vector; $\mathbf{f}(t) \in \mathbb{R}^{n \times 1}$ is the excitation force vector; $\mathbf{f}_c(\mathbf{q}, \dot{\mathbf{q}}, t) \in \mathbb{R}^{n \times 1}$ is nonlinear contact force interaction vector obtained after performing the Craig-Bampton reduction method.

To solve the equation (4) for periodic excitation using HBM ([32]), the periodic quantities (i.e. displacements and forces) with an angular frequency of $\omega$ are expressed as truncated series of harmonic terms:

$$\mathbf{q}(t) = \Re\left(\sum_{h=0}^{H} \widehat{\mathbf{q}}^{(h)} e^{ih\omega t}\right); \quad \mathbf{f}(t) = \Re\left(\sum_{h=0}^{H} \widehat{\mathbf{f}}^{(h)} e^{ih\omega t}\right); \quad \mathbf{f}_c(\mathbf{q},\dot{\mathbf{q}},t) = \Re\left(\sum_{h=0}^{H} \widehat{\mathbf{f}}_c^{(h)}(\widehat{\mathbf{q}}) e^{ih\omega t}\right) \quad (5)$$

where $\Re(\cdot)$ represents the real part of the quantity.

By applying Galerkin's procedure with the multi-harmonic approximation, the time domain nonlinear differential equation (4) is transformed into a nonlinear algebraic equation with Fourier coefficients as defined in equation (5), and the balance equation is written as:

$$\mathbf{d}^{(h)} \widehat{\mathbf{q}}^{(h)} = \widehat{\mathbf{f}}^{(h)} + \widehat{\mathbf{f}}_c^{(h)} \; with \; h = 0..H \quad (6)$$

where $\mathbf{d}^{(h)} = (-(h\omega)^2 \mathbf{m} + ih\omega \mathbf{c} + \mathbf{k})$ is the $h^{th}$ dynamic stiffness matrix of the system. The equation (6) consists the static ($h = 0$) and dynamic ($h = 1..H$) equations of the system coupled to each other by Fourier coefficients of the nonlinear contact force $\widehat{\mathbf{f}}_c$ depending on the Fourier coefficients of the displacement $\widehat{\mathbf{q}}$. The number of harmonics $H$ is chosen in a way to approximate the dynamics of the structure with sufficient accuracy. The residual equation is formulated by rewriting equation (6) as:

$$\mathbf{RES}^{(h)} = \mathbf{d}^{(h)} \widehat{\mathbf{q}}^{(h)} - \widehat{\mathbf{f}}^{(h)} - \widehat{\mathbf{f}}_c^{(h)} \; with \; h = 0..H \quad (7)$$

## 2.2 Governing equations of a shrouded bladed disk

Let us consider a shrouded bladed disk with shroud contact interface, as shown in Figure 1. The bladed disk is a tuned system because it is made of identical fundamental sectors arranged in a cyclical manner. Since the bladed disks are cyclic in nature, one can exploit the analysis by reducing the full bladed disk model to a single fundamental sector, by applying proper boundary conditions at the sector interfaces in the nonlinear dynamic analysis ([17], [34]).

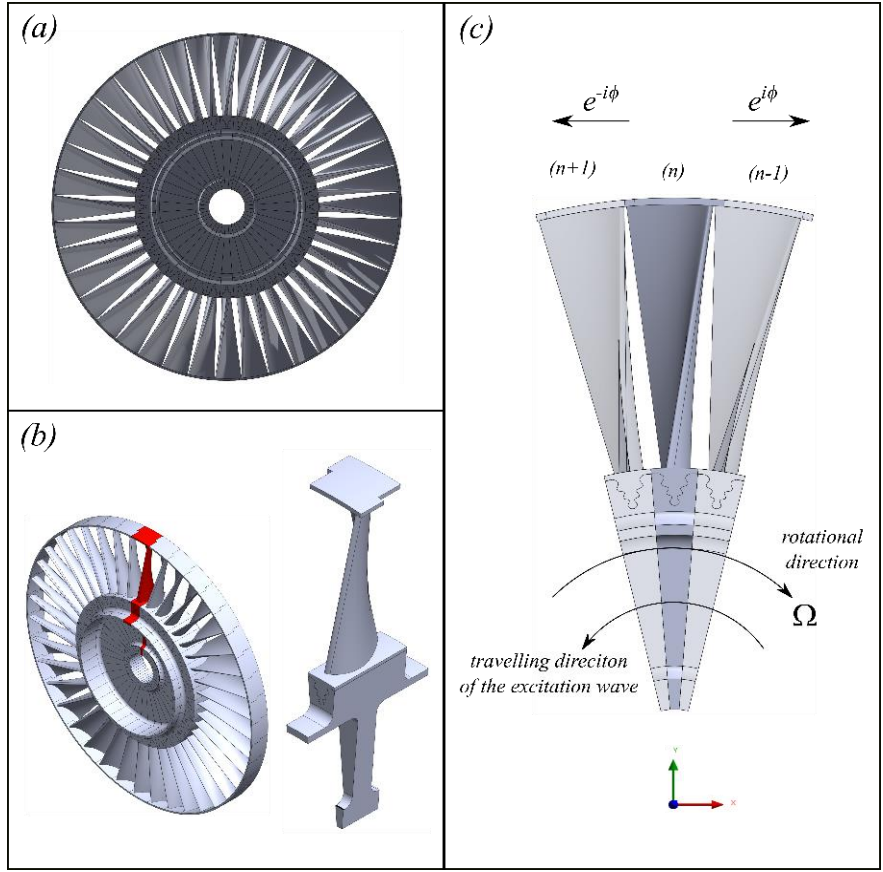

Figure 1: (a) Full bladed disk with shroud contacts (b) A fundamental blade-disk sector
(c) Schematic view of cyclic symmetry for bladed disk model

The equation of motion for a tuned bladed disk with contact interfaces with $N_s$ sectors can be obtained by modifying equation (1) as:

$$\mathbf{M}^{(n)}\ddot{\mathbf{Q}}(t) + \mathbf{C}^{(n)}\dot{\mathbf{Q}}(t) + \mathbf{K}^{(n)}\mathbf{Q}(t) = {}^{(n)}\mathbf{F}(t) + {}^{(n)}\mathbf{F}_{c,l}(\Delta\mathbf{q}_{c,l}, \Delta\dot{\mathbf{q}}_{c,l}, t) + {}^{(n)}\mathbf{F}_{c,r}(\Delta\mathbf{q}_{c,r}, \Delta\dot{\mathbf{q}}_{c,r}, t) \qquad (8)$$

where $n = 1..N_s$ is a sector number; $\mathbf{M} \in \mathbb{R}^{N\times N}, \mathbf{C} \in \mathbb{R}^{N\times N}$, and $\mathbf{K} \in \mathbb{R}^{N\times N}$ denote the mass matrix, viscous damping matrix and the stiffness matrix respectively of $n^{th}$ isolated bladed disk segment without cyclic boundary conditions; ${}^{(n)}\mathbf{Q}(t) \in \mathbb{R}^{N\times 1}$ is a nodal displacement vector; ${}^{(n)}\mathbf{F}(t) \in \mathbb{R}^{N\times 1}$ is the excitation force vector acting on $n^{th}$ blade, and ${}^{(n)}\mathbf{F}_{c,l}, {}^{(n)}\mathbf{F}_{c,r} \in \mathbb{R}^{N\times 1}$ are nonlinear contact force interaction vectors at the shrouds of the left and the right adjacent sectors respectively depending on the relative displacements and velocities. These contact forces depend on the relative displacements in a nonlinear fashion at the shroud contact.

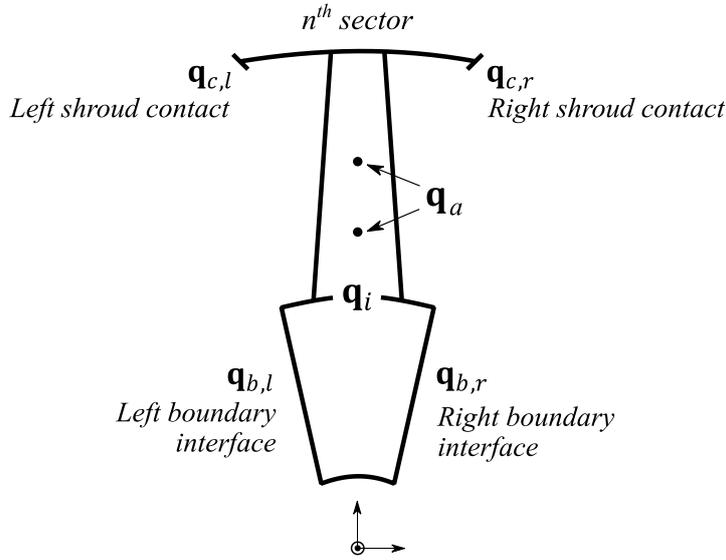

Figure 2: A schematic view of the isolated bladed disk segment

The displacement vector at $n^{th}$ isolated bladed disk segment is written as:

$$^{(n)}\mathbf{Q} = \begin{bmatrix} {}^{(n)}\mathbf{q}_{b,l}{}^T & {}^{(n)}\mathbf{q}_{b,r}{}^T & {}^{(n)}\mathbf{q}_a{}^T & {}^{(n)}\mathbf{q}_{c,l}{}^T & {}^{(n)}\mathbf{q}_{c,r}{}^T & {}^{(n)}\mathbf{q}_i{}^T \end{bmatrix}^T \qquad (9)$$

where ${}^{(n)}\mathbf{q}_{b,l}$ and ${}^{(n)}\mathbf{q}_{b,r}$ are the nodal displacements of the left boundary segment interface and right boundary segment interface, respectively, as shown in Figure 2. ${}^{(n)}\mathbf{q}_{c,l}$ and ${}^{(n)}\mathbf{q}_{c,r}$ are the displacements of left and right shroud contact nodes, respectively. ${}^{(n)}\mathbf{q}_a$ are the accessory nodes which contains the displacements of loading and response nodes, and ${}^{(n)}\mathbf{q}_i$ contains the displacements of other interior nodes of the system of the $n^{th}$ segment.

Similar to the generic system, to solve the equation (8) for periodic excitation using HBM ([32]), the periodic quantities (i.e. displacements and forces) with an angular frequency of $\omega$ are expressed as truncated series of harmonic terms:

$$^{(n)}\mathbf{Q}(t) = \Re\left(\sum_{h=0}^{H} {}^{(n)}\widehat{\mathbf{Q}}^{(h)} e^{ih\omega t}\right); \quad {}^{(n)}\mathbf{F}(t) = \Re\left(\sum_{h=0}^{H} {}^{(n)}\widehat{\mathbf{F}}^{(h)} e^{ih\omega t}\right); \qquad (10)$$

$$^{(n)}\mathbf{F}_{c,l}(\Delta \mathbf{q}_{c,l}, \Delta \dot{\mathbf{q}}_{c,l}, t) = \Re\left(\sum_{h=0}^{H} {}^{(n)}\hat{\mathbf{F}}_{c,l}^{(h)} e^{ih\omega t}\right); \quad {}^{(n)}\mathbf{F}_{c,r}(\Delta \mathbf{q}_{c,r}, \Delta \dot{\mathbf{q}}_{c,r}, t) = \Re\left(\sum_{h=0}^{H} {}^{(n)}\hat{\mathbf{F}}_{c,r}^{(h)} e^{ih\omega t}\right)$$

By applying Galerkin's procedure with the multi-harmonic approximation, the time domain nonlinear differential equation (8) is transformed into a nonlinear algebraic equation with Fourier coefficients as defined in equation (10), and the balance equation is written as:

$$\mathbf{D}^{(h)} {}^{(n)}\hat{\mathbf{Q}}^{(h)} = {}^{(n)}\hat{\mathbf{F}}^{(h)} + {}^{(n)}\hat{\mathbf{F}}_{c,l}^{(h)} + {}^{(n)}\hat{\mathbf{F}}_{c,r}^{(h)} \; with \; h = 0..H \tag{11}$$

where $\mathbf{D}^{(h)} = (-(h\omega)^2 \mathbf{M} + ih\omega \mathbf{C} + \mathbf{K})$ is the $h^{th}$ dynamic stiffness matrix of the system. The equation (11) consists of the static ($h=0$) and dynamic ($h=1..H$) equations of the system coupled to each other by Fourier coefficients of the nonlinear contact force ${}^{(n)}\hat{\mathbf{F}}_{c,l}^{(h)}$ and ${}^{(n)}\hat{\mathbf{F}}_{c,r}^{(h)}$ depending on the Fourier coefficients of the relative displacements and velocities. The number of harmonics $H$ is chosen in a way to approximate the steady state dynamics of the structure with sufficient accuracy. The equation (11) has to be solved iteratively for the unknown displacement vector of $n^{th}$ segment because of the nonlinear contact forces. Note the size of the dynamic stiffness matrix $\mathbf{D} \in \mathbb{C}^{(N(H+1) \times N(H+1))}$ and the force vectors ${}^{(n)}\hat{\mathbf{F}}, {}^{(n)}\hat{\mathbf{F}}_{c,l}, {}^{(n)}\hat{\mathbf{F}}_{c,r} \in \mathbb{C}^{(N(H+1) \times 1)}$. The system of equations still refer to the single fundamental bladed disk sector. The cyclic symmetry boundary conditions will be implemented in the next steps.

## 2.3 Implementation of cyclic symmetry constraints

The turbine bladed disk is assumed to be subjected to a travelling wave type excitation, see ref. [17], [34] for detailed explanation. The inter-blade phase angle (IBPA) is defined as: $\emptyset = \frac{2\pi}{N}.EO$, where $N$ is the number of blades and $EO$ is the engine order that, in the Campbell diagram, crosses the resonance under investigation. Reducing the number of DOFs required for the nonlinear forced response analysis is done in two steps. First, the disk interface segment boundary DOFs are reduced using cyclic symmetry constraints and then similar constraints are applied to the shroud contact node DOFs.

**Reducing the disk segment boundary DOFs using cyclic symmetry constraints:**

Since the bladed disk is subjected to the travelling wave type excitation, it is assumed that the steady-state vibration response also represents a travelling wave. This results in a relationship between the right segment interface nodal DOF, and the left segment interface nodal DOF. Owing to the cyclic symmetry conditions based on the assumption of the turbine bladed disk is made of the similar bladed-disk fundamental sector, one can write the following coupling relationship for the displacements at the segment boundary DOFs:

$$^{(n)}\hat{\mathbf{q}}_{b,l}^{(h)} = {}^{(n)}\hat{\mathbf{q}}_{b,r}^{(h)} e^{-ih\emptyset} \tag{12}$$

Redefining the reduced displacement vector of unknowns by considering the right-side boundary segment for $h^{th}$ harmonic component and $n^{th}$ sector as:

$$^{(n)}\hat{\mathbf{Q}}^{(h)} = \hat{\mathbf{T}}_{CS}^{(h)} {}^{(n)}_{red}\hat{\mathbf{Q}}^{(h)} \tag{13}$$

Thereby making the new reduced displacement vector of unknowns as:

$$^{(n)}_{red}\hat{\mathbf{Q}}^{(h)} = \left[ {}^{(n)}\hat{\mathbf{q}}_{b,r}^{(h)^T} \; {}^{(n)}\hat{\mathbf{q}}_{a}^{(h)^T} \; {}^{(n)}\hat{\mathbf{q}}_{c,l}^{(h)^T} \; {}^{(n)}\hat{\mathbf{q}}_{c,r}^{(h)^T} \; {}^{(n)}\hat{\mathbf{q}}_{i}^{(h)^T} \right]^T \tag{14}$$

$$\widehat{\mathbf{T}}_{CS}^{(h)} = \begin{bmatrix} \mathbf{I}(e^{-ih\emptyset}) & 0 & 0 & 0 & 0 \\ \mathbf{I} & 0 & 0 & 0 & 0 \\ 0 & \mathbf{I} & 0 & 0 & 0 \\ 0 & 0 & \mathbf{I} & 0 & 0 \\ 0 & 0 & 0 & \mathbf{I} & 0 \\ 0 & 0 & 0 & 0 & \mathbf{I} \end{bmatrix} \tag{15}$$

Where $\widehat{\mathbf{T}}_{CS}^{(h)}$ is the cyclic symmetry constraint matrix for $h^{th}$ harmonic component; $\mathbf{I}$ is the identity matrix of an appropriate size corresponding to the size of DOFs. Substituting the vector $^{(n)}\widehat{\mathbf{Q}}^{(h)}$ in equation (11) by the cyclic constraint relation as given in equation (13) for $h = 0..H$ and left-multiplying by the corresponding complex conjugate transpose of $\widehat{\mathbf{T}}_{CS}^{(h)}$ results in:

$$_{red}\widehat{\mathbf{D}}^{(h)}\,_{red}^{(n)}\widehat{\mathbf{Q}}^{(h)} = \,_{red}^{(n)}\widehat{\mathbf{F}}^{(h)} + \,_{red}^{(n)}\widehat{\mathbf{F}}_{c,l}^{(h)} + \,_{red}^{(n)}\widehat{\mathbf{F}}_{c,r}^{(h)} \text{ with } h = 0..H \tag{16}$$

where $_{red}\widehat{\mathbf{D}}^{(h)} = \left(-(h\omega)^2\widehat{\mathbf{M}}^{(h)} + ih\omega\widehat{\mathbf{C}}^{(h)} + \widehat{\mathbf{K}}^{(h)}\right)$ is the $h^{th}$ dynamic stiffness matrix of the system. Note that the complex mass matrix $\widehat{\mathbf{M}} \in \mathbb{C}^{(j(H+1) \times j(H+1))}$, viscous damping matrix $\widehat{\mathbf{C}} \in \mathbb{C}^{(j(H+1) \times j(H+1))}$, and the stiffness matrix $\widehat{\mathbf{K}} \in \mathbb{C}^{(j(H+1) \times j(H+1))}$ are Hermitian because of the application of the cyclic symmetry constraints and $_{red}^{(n)}\widehat{\mathbf{F}}, \,_{red}^{(n)}\widehat{\mathbf{F}}_{c,l}, \,_{red}^{(n)}\widehat{\mathbf{F}}_{c,r} \in \mathbb{C}^{(j(H+1) \times 1)}$ where $j(H+1) < N(H+1)$ because of the successful elimination of the left boundary segment interface DOFs.

**Reducing the shroud contact DOFs using cyclic symmetry constraints:**

The displacement vector of unknowns can further be reduced by applying the cyclic symmetry constraints at the shroud contact DOFs and writing in relative displacement terms. Writing in relative displacements terms reduces the number of shroud contact DOFs by half as one can consider only the number of shroud contact DOFs of one side and reduce the size of the system required for nonlinear analysis. The relative displacement at the right-side shroud contact and the non-linear contact force can be written as:

$$^{(n)}\Delta\widehat{\mathbf{q}}_{c,r}^{(h)} = \,^{(n)}\widehat{\mathbf{q}}_{c,r}^{(h)} - \,^{(n-1)}\widehat{\mathbf{q}}_{c,l}^{(h)}; \,^{(n)}\Delta\widehat{\mathbf{q}}_{c,r}^{(h)} = \,^{(n)}\widehat{\mathbf{q}}_{c,r}^{(h)} - \,^{(n)}\widehat{\mathbf{q}}_{c,l}^{(h)}e^{ih\emptyset}$$
$$^{(n)}\widehat{\mathbf{f}}_{c,r}^{(h)} = -\,^{(n)}\widehat{\mathbf{f}}_{c,l}^{(h)}e^{ih\emptyset} \tag{17}$$

The new displacement vector of unknowns in terms of relative displacements at the shroud contact becomes:

$$_{rel}^{(n)}\widehat{\mathbf{Q}}^{(h)} = \left[^{(n)}\widehat{\mathbf{q}}_{b,r}^{(h)T} \,^{(n)}\widehat{\mathbf{q}}_a^{(h)T} \,^{(n)}\Delta\widehat{\mathbf{q}}_{c,r}^{(h)T} \,^{(n)}\widehat{\mathbf{q}}_i^{(h)T}\right]^T \tag{18}$$

This leads to cutting down the shroud contact unknown DOFs by half by considering either right or left contact, hence reducing the size of the non-linear computation. Similarly, a relative displacement cyclic symmetry transformation matrix [35] can be defined, and the balance equation (16) can be rewritten as:

$$_{rel}\widehat{\mathbf{D}}^{(h)}\,_{rel}^{(n)}\widehat{\mathbf{Q}}^{(h)} = \,_{rel}^{(n)}\widehat{\mathbf{F}}^{(h)} + \,_{rel}^{(n)}\widehat{\mathbf{F}}_{c,r}^{(h)} \text{ with } h = 0..H \tag{19}$$

where $\widehat{\mathbf{M}}, \widehat{\mathbf{C}}, \widehat{\mathbf{K}} \in \mathbb{C}^{(k(H+1) \times k(H+1))}$ and $_{rel}^{(n)}\widehat{\mathbf{F}}, \,_{rel}^{(n)}\widehat{\mathbf{F}}_{c,r} \in \mathbb{C}^{(k(H+1) \times 1)}$ where $k(H+1) < j(H+1)$.

Since the number of shroud contact nodes, boundary nodes and accessory nodes DOFs are much smaller than the number of internal DOFs, as described for the generic system, the classical Craig-Bampton reduction method [33] is applied, retaining as physical coordinates the right segment interface nodes, right shroud contact nodes, accessory nodes DOFs – collectively called as master nodes, and adding a certain number of fixed-interface modes $\boldsymbol{\eta}$.

$$^{(n)}\widehat{\mathbf{q}}^{(h)} = \left[^{(n)}\widehat{\mathbf{q}}_{b,r}^{(h)T} \,^{(n)}\widehat{\mathbf{q}}_a^{(h)T} \,^{(n)}\Delta\widehat{\mathbf{q}}_{c,r}^{(h)T} \,^{(n)}\boldsymbol{\eta}^T\right]^T \tag{20}$$

where $^{(n)}\hat{\mathbf{q}} \in \mathbb{C}^{m(H+1)\times 1}$ and $m(H+1) \ll k(H+1)$. Finally, the equation (19) can be reformulated using equation (20) as:

$$\hat{\mathbf{d}}^{(h)}\,^{(n)}\hat{\mathbf{q}}^{(h)} = \,^{(n)}\hat{\mathbf{f}}^{(h)} + \,^{(n)}\hat{\mathbf{f}}_{c,r}^{(h)} \quad with\ h = 0..H \tag{21}$$

By rearranging equation (21), the residual equation is written as:

$$\mathbf{RES}^{(h)} = \hat{\mathbf{d}}^{(h)}\,^{(n)}\hat{\mathbf{q}}^{(h)} - \,^{(n)}\hat{\mathbf{f}}^{(h)} - \,^{(n)}\hat{\mathbf{f}}_{c,r}^{(h)} \quad with\ h = 0..H \tag{22}$$

To use the nonlinear solvers, the residual has to be supplied in the real form. Hence, the complex form residual equation (7) and equation (22) is split into its real and imaginary components to real form to minimize the residual to an acceptable tolerance and find the unknown displacement vector as:

$$\mathbf{RES} = \left[\mathbf{RES}^{(0)}, \Re(\mathbf{RES}^{(1)}), \Im(\mathbf{RES}^{(1)}) \ldots \Re(\mathbf{RES}^{(H)}), \Im(\mathbf{RES}^{(H)})\right]^{\mathrm{T}} \tag{23}$$

## 2.4 Contact Model

A contact model is necessary to compute the nonlinear contact forces mentioned in the previous section. There are various node-to-node and patch-to-patch contact models available in the literature ([36], [37], [46]–[48], [38]–[45]) such as Coulomb slider, Jenkins element – 2D and 3D with constant and variable normal load, Iwan model and its variations, Bouc-Wen model, Valanis model, LuGre model etc. In our study, we chose the state-of-the-art node-to-node 2D Jenkins element with a variable normal load to compute the contact forces, as shown in Figure 3(a). A typical hysteresis loop generated for the given input force and displacements is shown in Figure 3(b). The solution to the balance equations requires the contact forces on the contact interface as input. The contact element is characterized by two linear springs in the tangential and normal direction with tangential ($k_t$) and normal ($k_n$) contact stiffness, at each node pair over the contact interface. The contact model allows to characterize and simulate three possible contact states – stick, slip and lift-off.

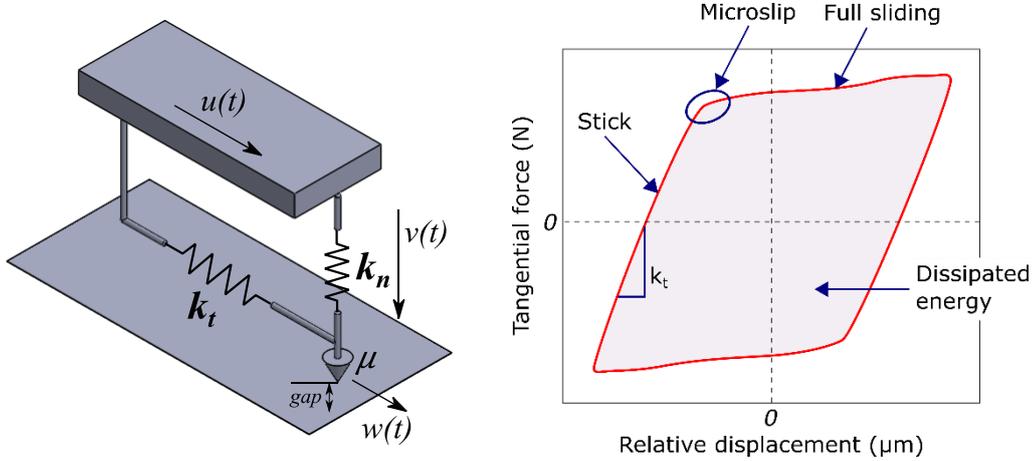

Figure 3: (a) Jenkin's element contact model with variable normal load, (b) A typical hysteresis loop

At each element, the contact forces – tangential force $T(t)$ and the normal force $N(t)$ depend on the relative tangential and normal displacements of the corresponding node pair, namely $\mathbf{u}(t)$ and $\mathbf{v}(t)$ respectively [29]. The friction limit value is defined by the Coulomb law as $\mu N(t)$, where $\mu$ is the coefficient of friction. To consider the friction phenomena occurring at the interface in the tangential direction, a slider is used to connect the two bodies. When the tangential force exceeds the limit value, the slider starts moving, and the amount of slip between the nodes is $\mathbf{w}(t)$.

Since the contact is a unilateral constraint, the normal contact force $N(t)$ is defined at each time $t$ as:

$$N(t) = \max(k_n \mathbf{v}(t), 0) \tag{24}$$

When $\mathbf{v}(t)$ is negative, no normal contact force is allowed, and separation occurs, hence referred to as lift-off state.

For the tangential direction, the tangential contact force $T(t)$ is dependent on the contact state as defined by:

$$T(t) = \begin{cases} k_t\{\mathbf{u}(t) - \mathbf{w}(t)\} & \text{stick state} \\ \mu N(t) sign(\dot{\mathbf{w}}) & \text{slip state} \\ 0 & \text{lift-off state} \end{cases} \quad (25)$$

The contact forces are computed using predictor-corrector logic [17]. At each time $t$ a predictor step is performed assuming the stick contact conditions:

$$T^p(t) = k_t[\mathbf{u}(t) - \mathbf{w}(t)] = k_t[\mathbf{u}(t) - \mathbf{w}(t - \Delta t)] \quad (26)$$

Where $\Delta t$ is the time step, then a corrector step is performed to compute the actual value of $T(t)$ given by:

$$T(t) = \begin{cases} T^p(t) & \text{stick state} \\ \mu N(t) sign(T^p(t)) & \text{slip state} \\ 0 & \text{lift-off state} \end{cases} \quad (27)$$

The slider displacement $\mathbf{w}(t)$ is computed as:

$$\mathbf{w}(t) = \begin{cases} \mathbf{w}(t - \Delta t) & \text{stick state} \\ \mathbf{u}(t) - \mu N(t) sign(T(t))/k_t & \text{slip state} \\ \mathbf{u}(t) & \text{lift-off state} \end{cases} \quad (28)$$

The approximation introduced due to the predictor-corrector method is minimized by choosing a fine time step $\Delta t$ (for instance, 2e8 time steps).

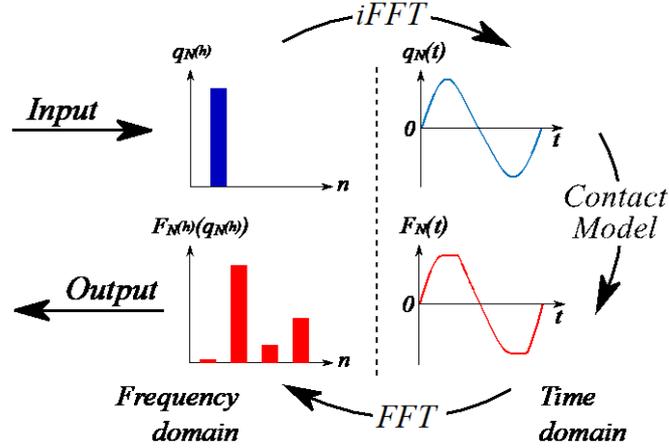

Figure 4: Alternating Frequency Time (AFT) method

The balance equations are solved in the frequency domain using HBM, whereas the accurate contact forces are possible to compute only in the time domain. Hence an Alternate Frequency/Time (AFT) method ([49], [50]) is employed. The relative displacements are converted from the frequency domain to the time domain by applying Inverse Fast Fourier Transform and then run through the contact model. The contact forces are obtained in the time domain. Then applying Fast Fourier Transform to convert to the Fourier coefficient of contact forces in the time domain, as shown in Figure 4.

Newton-Raphson logic is implemented for the iterative procedure. The Jacobian matrix needed to compute the residual for the Newton-Raphson method is computationally intensive if using MATLAB built-in finite difference method (FDM) for large systems. Hence an analytical Jacobian is implemented based on the works of [17], [51]. The contact forces and the partial derivatives of the contact forces are computed and assembled in the Jacobian matrix. This dramatically increases the speed of the solution by many folds. The contact forces and computation of the dynamic response are primarily based on the accurate input of the normal and tangential contact stiffness. The closed-form solutions needed to compute the accurate tangential ($k_t$) and normal ($k_n$) contact stiffness to provide the contact model is provided based on the analytical equations as mentioned in Ref [52]–[55]. Then the contact stiffnesses are distributed over the contact area according to the individual elemental contribution.

## 2.5 Wear computation

To predict the effect of fretting wear on the dynamic forced response of the bladed disk, wear has to be evaluated. There are many wear models [56] available in the literature, such as Archard's law [2], wear energy approach [31], thermodynamics approach, etc. Due to its simple nature in the formulation, good correlation with experiments, lower computation effort and easy implementation in the forced response solvers to evaluate the non-linear dynamics, a wear energy approach is used. Many research studies have shown the wear energy approach in good agreement with the experimental results ([6], [7], [57]). Initially, the dynamic forced response is computed with pristine contact surfaces without any wear. The nodal wear depth $\mathbf{v}_w$ is defined as:

$$\mathbf{v}_w = Z_W \frac{\alpha E}{A} \qquad (29)$$

where $A$ is the area associated with the node pair, $\alpha$ is the wear coefficient for a particular contact pair, and $E$ is the energy dissipated over one cycle. It is worth mentioning that $A$ depends on the finite element mesh over the contact surface, while $\alpha$ is the experimentally obtained parameter for a specific material couple ([58]–[60]) and $E$ is obtained by evaluating the area under the hysteresis loop. The control parameter of equation (29) is $Z_W$ – number of cycles.

The process of updating the contact surface due to wear after one cycle is called a wear iteration. With materials typical of turbine blades, the wear depth obtained at each vibration cycle is so small that no significant effect on the forced response of the system is observed if a wear iteration is performed at each vibration cycle. Hence, a wear acceleration technique is used, where the wear depth $\mathbf{v}_w$ associated at each node pair is considered for $Z_W$ number of cycles.

In this paper, an adaptive strategy is used, and the value of $Z_W$ at each wear iteration is computed as:

$$Z_W = floor\left(\frac{\mathbf{v}_{w,max}}{max(\Delta h_{ij})}\right) \qquad (30)$$

where, $\mathbf{v}_{w,max}$ is a user-defined parameter and $\Delta h_{ij}$ is the nodal wear depth at the contact patch for one vibration cycle. Choosing $\mathbf{v}_{w,max}$ is empirical at this stage, as there are no guidelines from the experiments what value is most suitable in terms of accuracy of results. However, we present a few thumb rules one can use to define the parameter $\mathbf{v}_{w,max}$:

- the maximum wear depth allowed for each wear iteration, directly selected by the user

    For example, let us consider an assembly with a contact interface is functional up to a contact wear depth of 100 µm. In this case, the user can arbitrarily discretize this maximum allowable wear depth into 100 parts and define $\mathbf{v}_{w,max}$ as 1 µm. By doing this, the user can visualise the contact evolution and the effect on the non-linear response plots at sufficient intervals before the functional failure of the assembly.

- the percentage of the maximum static deflection for the given static loads acting at the contact

    This method is more quantitative than the previous arbitrary method. $\mathbf{v}_{w,max}$ is a parameter of the static loading condition. Irrespective of the geometry, size and loading scenario of the system, $\mathbf{v}_{w,max}$ can be defined as a percentage of the maximum static deflection at the contact. For example, $\mathbf{v}_{w,max}$ can be chosen as 0.1% to 10% of the maximum static deflection ($\delta_{max}$).

- the maximum tolerable error in wear depth

  This is more of a special case. Let us consider that the user has a very strict requirement of the accuracy of wear depth needed in each wear iteration. Then $\mathbf{v}_{w,max}$ can be defined directly as the value of this wear depth.

In this study, we define $\mathbf{v}_{w,max}$ as the percentage of the maximum static deflection at the contact patch for the given static loading condition. This percentage is varied between 0.5% to 10% of the maximum static deflection ($\delta_{max}$). This parameter determines $Z_W$, in turn, is a trade-off between the accuracy of the results and computation time needed. A low value of $Z_W$ will allow for a more accurate, but more time-consuming analysis, the opposite will happen if a high value of $Z_W$ is selected.

Once the term $\mathbf{v}_w$ is computed using equation (29), the wear iteration is performed by updating the 0$^{th}$ order Fourier coefficient of the normal relative displacement at the contact in the following way:

$$\mathbf{v}(t) = \sum_{h=0}^{H} \hat{\mathbf{v}}^{(h)} e^{ih\omega t} - \mathbf{v}_w \tag{31}$$

After the wear iteration, the forced response of the updated system is computed by solving the updated balance equations, taking into account the effect of the wear depth on the static pre-load and the non-linear dynamics of the system simultaneously.

It is here worth mentioning that equation (31) allows for a direct update of the governing equations of the system since the static term (h = 0) is included in the set of non-linear equations to solve. This method allows computing the effect of wear on the forced response with 'changing pre-load' without the need for a separate non-linear static analysis routine to compute the contact pre-load distribution.

The results are shown and discussed in the next section. Figure 5 gives an overview of the steps with pre-processing and solution phase to compute the effect of wear on the non-linear dynamic response. The pre-processing can be done using any standard commercial software such as ANSYS. In our case, ANSYS V19.0 is used to generate the FE model, perform CB-CMS and to extract the reduced Mass and Stiffness matrices. For the solution phase MATLAB R2018 is used.

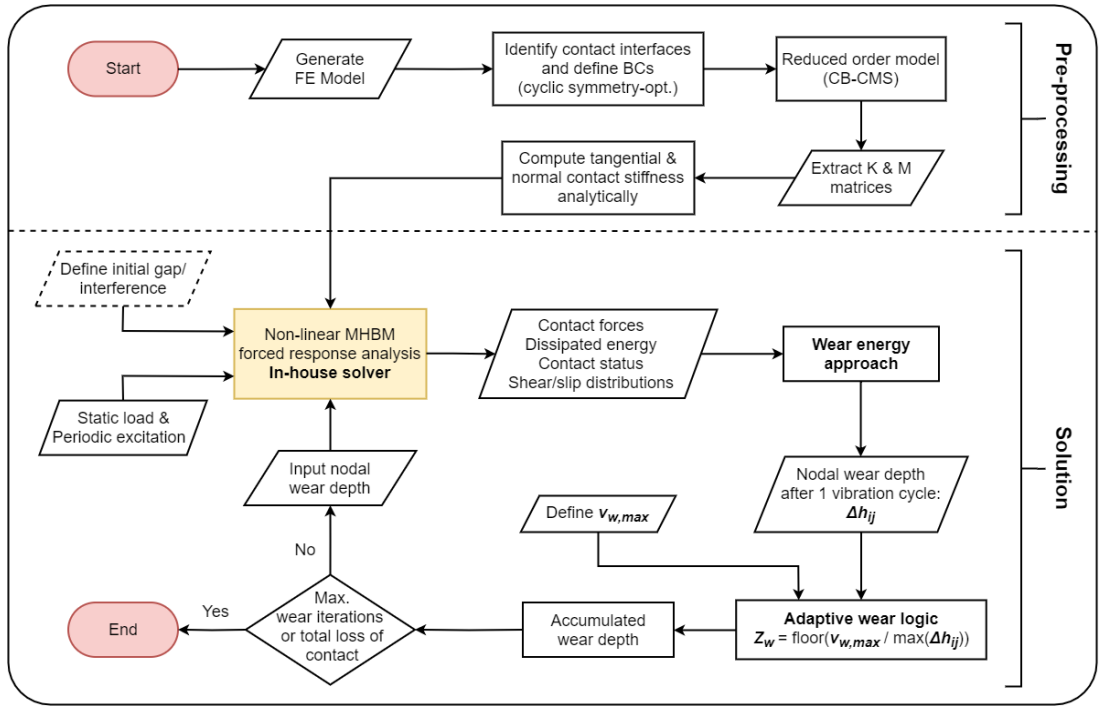

Figure 5: Flowchart showing an overview of steps to obtain the non-linear dynamic response with wear

# 3. Test Case – description, results and discussion

The main test case here is the tuned shrouded bladed disk with shroud contacts with cyclic symmetry boundary conditions. Before jumping on to the test case, the proposed method is first evaluated on a simple cantilever beam with a contact patch on one side and clamped at the other end. The two test cases demonstrate the proposed methodology with changing pre-load and the contact surface evolution during wear. The goal for the test cases is to choose the static load and the excitation in such a way that there is accelerated wear and total loss of contact due to wear with high energy dissipation at each cycle. This will aid validation in terms of the number of wear iterations, wear profile evolution and correlate the wear profile with the analytical free case. Also, the effect of the choice of parameter $\mathbf{v}_{w,max}$ is quantified in terms of accuracy and computational demand. In the following test cases, $\mathbf{v}_{w,max}$ is defined as a percentage of maximum static deflection under static loading conditions. Hence, a preliminary static analysis is performed to compute the maximum static deflection as if the contact was open without the presence of the second body.

## 3.1 Test Case 1: Cantilever beam

A steel cantilever beam with a contact patch and loading conditions as shown in Figure 6(a) and (b) and Table 1 is chosen as a test case to demonstrate the impact of wear on the dynamic response. The FE model mesh is shown in Figure 6(c). The contact elements at the contact patch are connected as node-to-node pair to the ground with the 2D Jenkins element contact model with the variable normal load, as shown in Figure 6(d). The beam is loaded at the mid-section so that as the wear progresses, the contact pre-load and the contact zone changes considerably. The contact patch is made up of 169 node pairs, and the grid distribution is as shown in Figure 7 with a response node as the centre node of the contact patch. Maximum static deflection ($\delta_{max}$) in this case is considered as the maximum deflection at the contact patch for the applied static load, as shown in Table 1.

Table 1: Parameters for the Cantilever beam and the contact patch

| Parameter | Value |
| --- | --- |
| Material | Steel |
| Young's modulus ($E$) | 210 GPa |
| Density ($\rho$) | 7860 kg/m$^3$ |
| Beam dimension | 200mm L x 20 mm H x 20mm D |
| Contact patch dimension | 20mm x 20mm |
| No. of contact elements | 169 |
| Tangential contact stiffness ($k_t$) | 470 N/µm |
| Normal contact stiffness ($k_n$) | 583 N/µm |
| Friction coefficient ($\mu$) | 0.5 |
| Wear energy coefficient ($\alpha$) | 2e3 µm$^3$/J |
| Static load ($F_{static}$) | 20 kN |
| Static load/Excitation ratio ($F_{static}/F_{ex}$) | 2 |
| Choice of $\mathbf{v}_{w,max}$ | {0.5%, 1%, 2%, 5%, 10%} of $\delta_{max}$ |

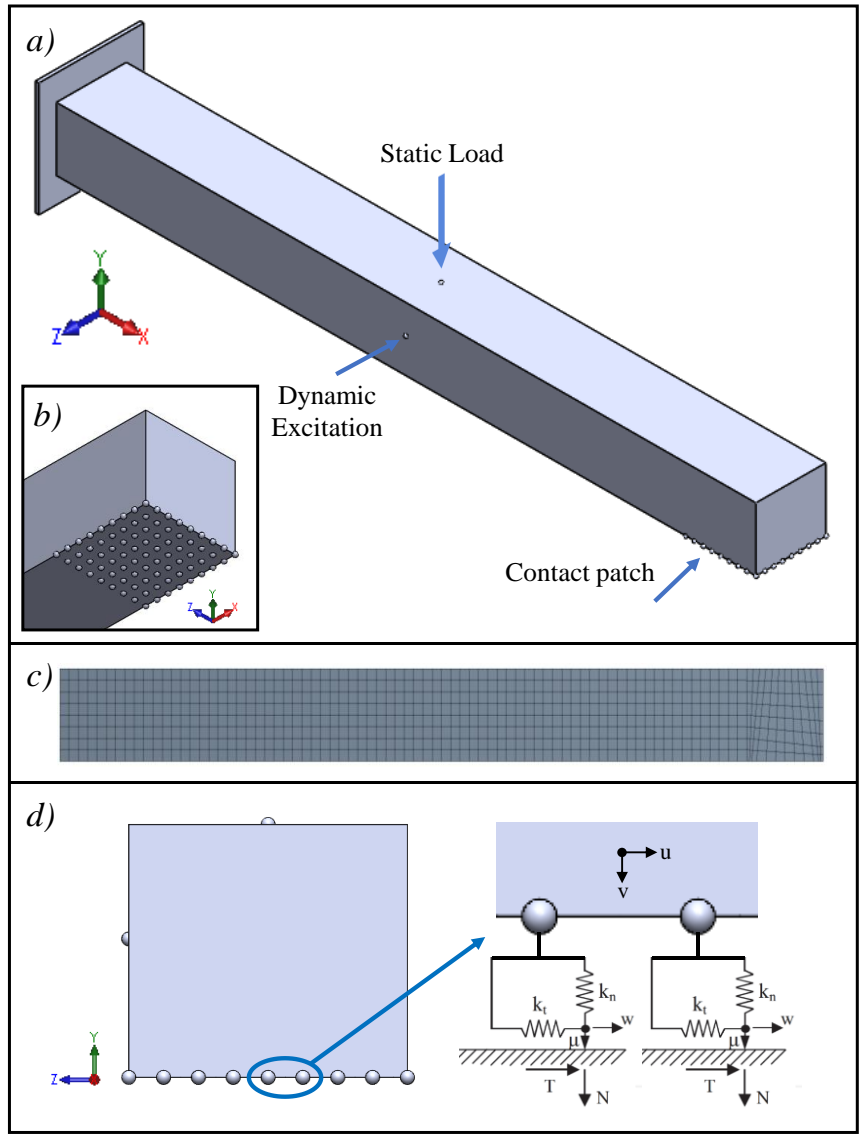

Figure 6: (a) 3D Model of a cantilever beam, (b) a close-up view of the contact patch, (c) FE mesh of the beam and (d) contact elements connection

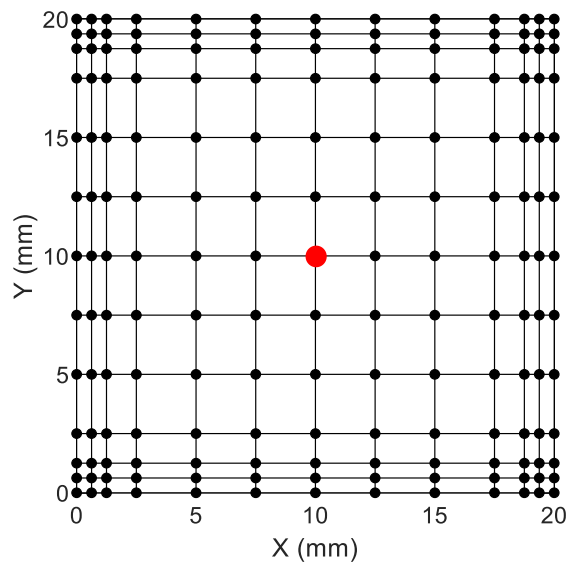

Figure 7: Contact patch with 169 elements and the centre response node (red dot)

The beam is excited by a harmonic transverse force, as shown in Figure 8, while the contact pre-load is provided by a vertical static load. Figure 8 also shows the static deflection and the shape at the initial contact condition with the second body and at the free condition when the second body is absent.

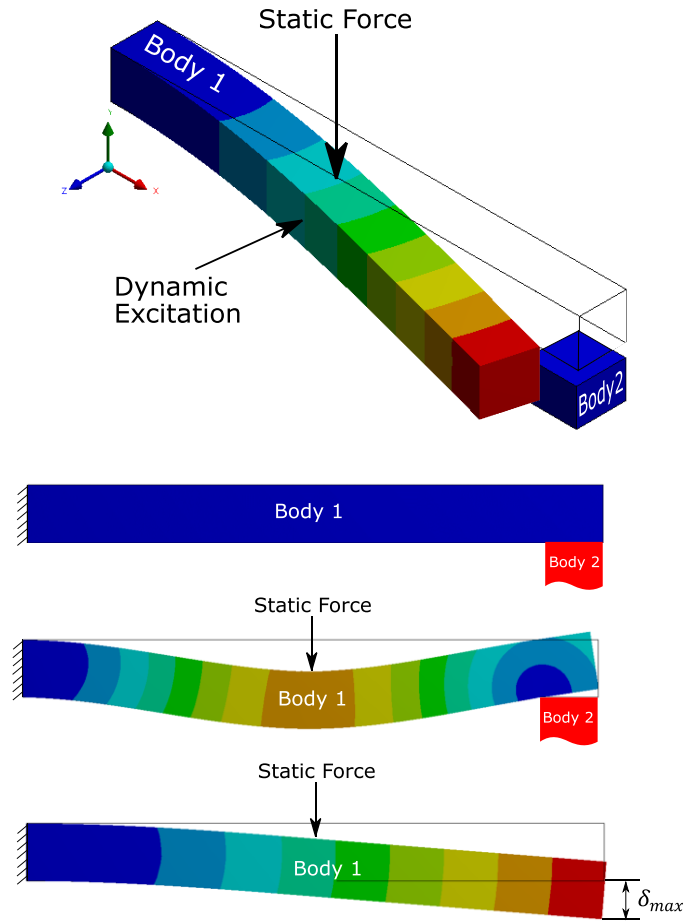

Figure 8: Excitation mode (first bending mode at 406 Hz) (top) and highlighting static deflection state with and without body 2 (bottom)

The FE model is generated using ANSYS, and the contact interface and boundary conditions are identified and defined in ANSYS. A Craig-Bampton Component Mode Synthesis (CB-CMS) [33] is performed by defining contact patch nodes and loading nodes as master nodes and retaining first 15 fixed interface modes to reduce the size of the system needed for the non-linear dynamic analysis. Otherwise, the dynamic analysis with all DOFs would be large and prohibitively time-consuming. Reduced mass and stiffness matrices are extracted after the reduction process. The reduced system has 534 DOFs. The solution algorithm is followed, as shown in Figure 5. $0^{th}$ and $1^{st}$ order harmonics are used to solve the static and dynamic balance equations in the frequency domain using HBM. Analytically computed tangential and normal contact stiffness is computed, taking into account the dimensions of the contact patch. These stiffnesses are distributed over the contact patch according to the elemental area contribution.

A preliminary linear analysis is performed to obtain free and stick states of the contact, and the responses at these states are assumed as reference. The computed non-linear response can then be compared with these two reference states. An initial gap/interference, if any, can also be defined using the current algorithm by offsetting in the direction normal to the contact. In the present case, the gap is set to 0 as the beam is resting on top of the body 2 before the static forces are applied to the system. Then both static and periodic forces are applied at the indicated excitation nodes. The contact forces, dissipated energy, contact status of each node and the shear/slip distributions are computed by solving equation (6). The wear energy approach, described in Section 2.5, is implemented to calculate the nodal wear depth at the contact patch for one vibration cycle. As the nodal wear depth is so minuscule at each vibration cycle, the dynamics is hardly affected. An adaptive wear acceleration logic is implemented by defining a parameter $v_{w,max}$ as described in the previous section. The parameter $v_{w,max}$ reduces the number of wear iterations, thus reducing the computation time, compared to running dynamic analysis for each vibration cycle. Since $v_{w,max}$ is a user-defined value, the effect of the choice of $v_{w,max}$ is also quantified. The iterative loop is performed until the required number of vibration cycles are obtained or until the complete loss of contact. Figure 9 shows the fineness of cumulative wear depth plots at the contact patch for different values of $v_{w,max}$.

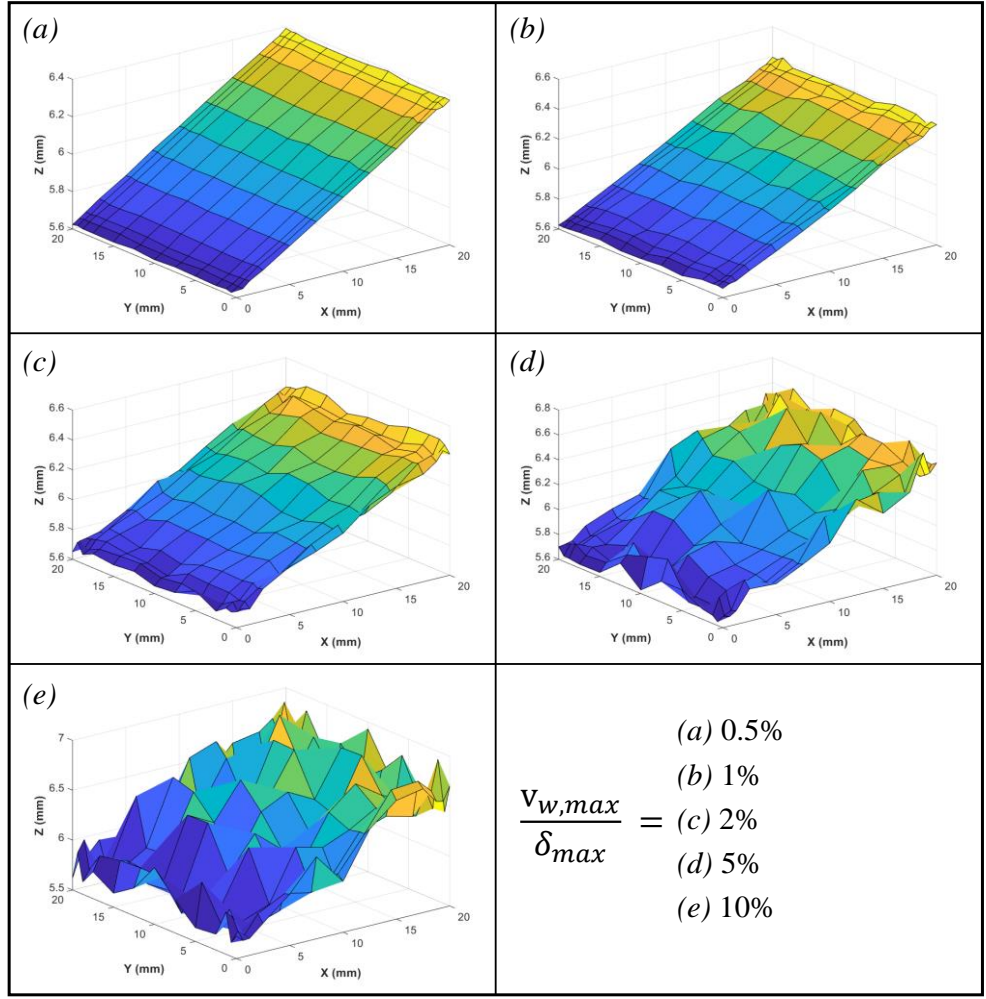

Figure 9: Cumulative wear depth plots at the contact patch for complete loss of contact for different values of $v_{w,max}$

Figure 10 shows the response curves of reference free and stick state along with the backbone curve of the maximum amplitude of response at each wear iteration for $v_{w,max}$ = 5% of $\delta_{max}$. It is evident and logical that, as the wear progresses, the contact interface loosens, and the contact tends towards the free state. The current formulation automatically takes into consideration the changing pre-load and is reflected in the response plots. Figure 11 provides an insight into the physical state of the system, cumulative wear depth and the contact status at different wear iterations. The wear iterations are chosen at the beginning, intermediate and at the end of the wear just before the complete loss of contact.

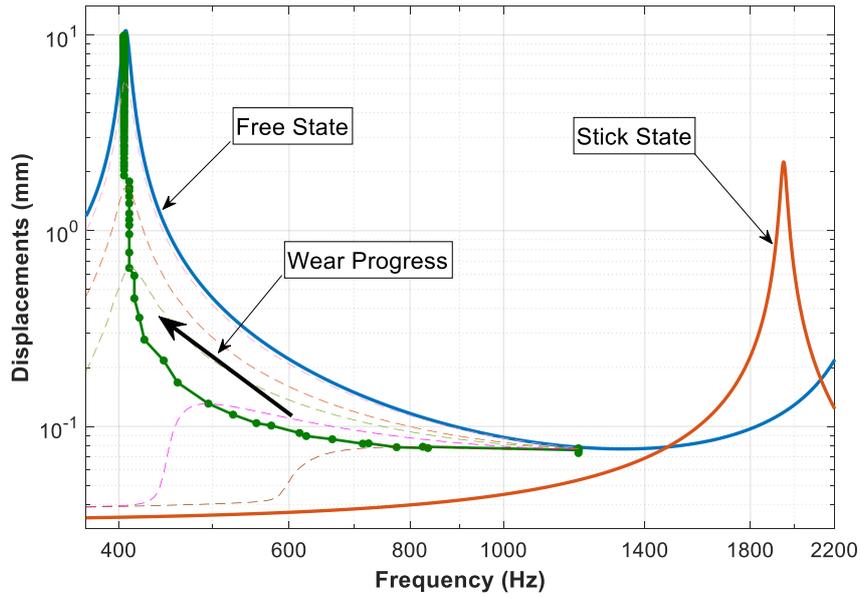

Figure 10: Response plot showing the backbone of the non-linear response obtained as wear progresses around first bending mode with reference free and stick states for $v_{w,max}$ = 5% of $\delta_{max}$

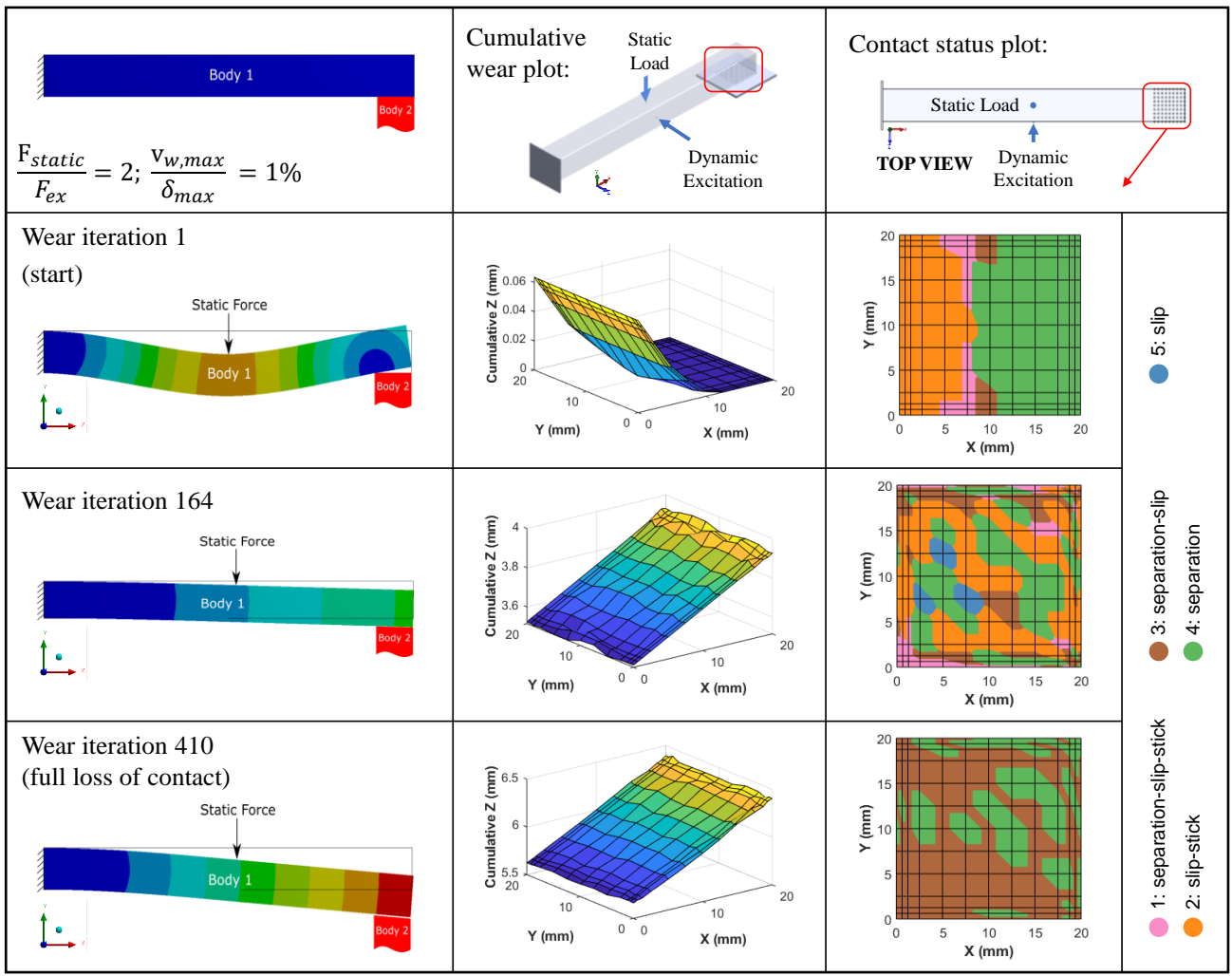

Figure 11: A result matrix showing - starting, intermediate and ending wear iteration just before the full loss of contact for the given test case highlighting the beam physical state, cumulative wear depth until that particular wear iteration and the contact status at that wear iteration

Figure 12 shows the plot of the number of vibration cycles at each wear iteration versus the number of wear iterations. For the shown plot, it took 183 wear iterations with the choice of $\mathbf{v}_{w,max}$ = 5% of $\delta_{max}$. In other words, with the allowed wear depth of 5% of maximum static deflection at each wear iteration, it took 183 wear iterations to completely lose the contact with the second body. Depending on the loading configuration, the shape of this graph can be different. In this example, as shown in Figure 11 of wear iteration 1, the contact started with a line contact at the inner edge of the contact between the body 1 and body 2, and proceeded to establish an area contact with an increase in wear iterations. The close-up of the contact state at the beginning and at the end, is also shown in Figure 12. The criteria to complete one wear iteration is the amount of wear depth achieved, which is predetermined by the user input. At the start of the wear and towards the end, a relatively high number of cycles are needed to achieve this limit of wear depth to satisfy the wear iteration because of partial contact. However, when the contact is fully engaged, lower number of cycles can satisfy the wear depth criteria because of large contact, hence more energy dissipation.

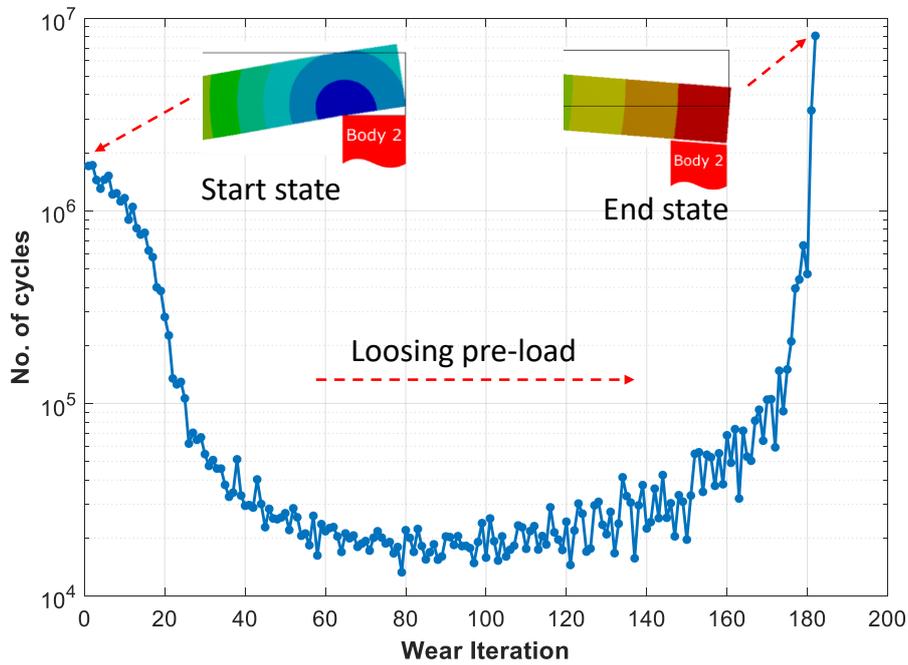

Figure 12: Wear iteration versus the number of vibration cycles plot showing the trend as the wear progresses from the start state to the complete loss of contact for $\mathbf{v}_{w,max}$ = 5% of $\boldsymbol{\delta_{max}}$
[Note: the number of cycles is adaptive at each wear iteration]

Figure 13 shows the maximum response amplitude at a particular wear iteration versus the cumulative number of cycles. It is evident the effect of $\mathbf{v}_{w,max}$ on the number of cycles at which the maximum response starts increasing. The accuracy of the proposed method for different values of $v_{w,max}$ is here investigated, by using as a reference the analysis performed with $\mathbf{v}_{w,max}/\delta_{max}$ = 0.5%, since no significant effects was observed by further reducing the value of $v_{w,max}$. Ideally, the best benchmark would be the response computed with $Z_W$ = 1 (one wear iteration per vibration cycle), but it is prohibitively time-consuming. It is clear that the choice of $\mathbf{v}_{w,max}$ is the trade-off between the computational cost and the accuracy of results; the lower the value of $\mathbf{v}_{w,max}$, the smoother the wear evolution and the contact patch wear profile is, due to the higher number of wear iterations. In general, a reasonable strategy should be based on a convergence analysis of the response as the value of $Z_W$ is progressively reduced. In this particular case, for relatively high values of $\mathbf{v}_{w,max}$ (from 2% to 10% of $\delta_{max}$) the analysis predicts the rapid increase in maximum response to occur earlier than the benchmark. Interestingly, when $\mathbf{v}_{w,max}/\delta_{max}$ = 1%, opposite results are obtained, showing that non-monotonic trend of the response during a convergence analysis can be observed.

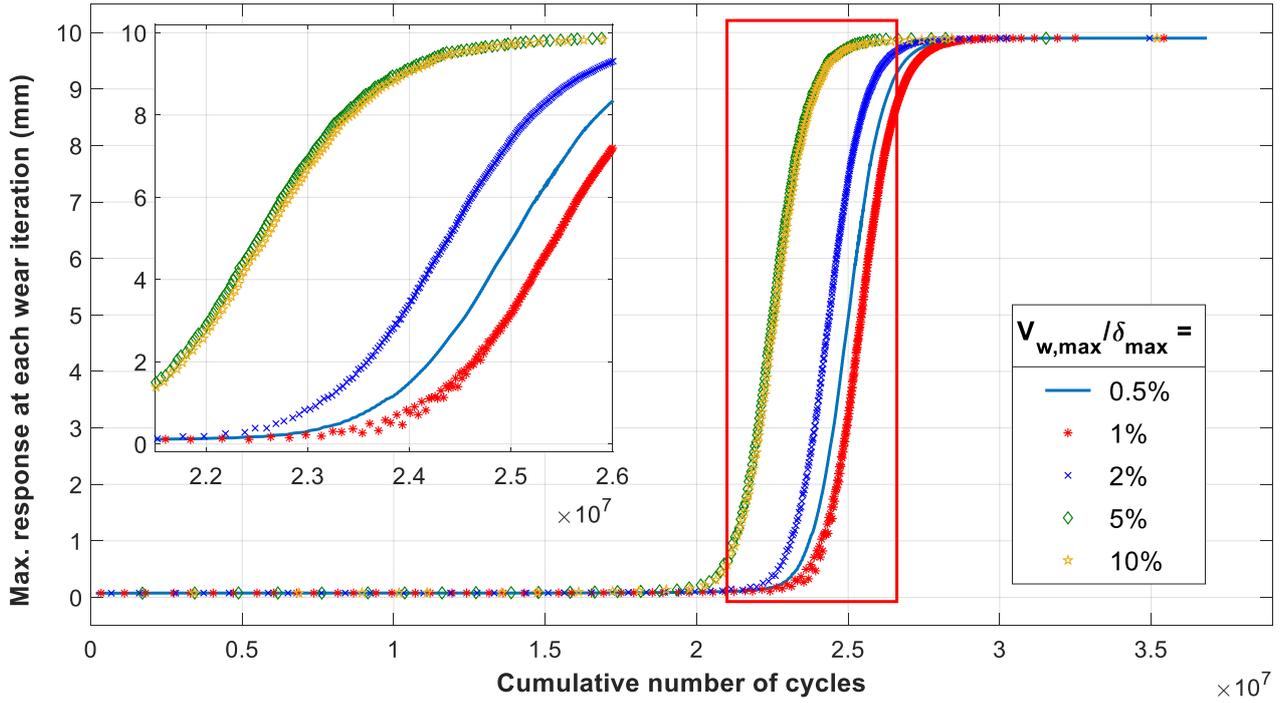

Figure 13: Maximum response at each wear iteration vs the cumulative number of cycles for different values of $\mathbf{v}_{w,max}$

Table 2 summarizes the comparison of the effect of the choice of $\mathbf{v}_{w,max}$ on the computational time, the number of wear iterations necessary for the complete loss of contact, maximum wear depth and the error in the maximum wear depth taking $\mathbf{v}_{w,max}$ = 0.5% as the reference. The trade-off effect of $\mathbf{v}_{w,max}$ is evident by comparing computational time vs error in wear depth.

Table 2: Comparison of the impact of wear analysis with non-linear dynamic response solver for different values of $\mathbf{v}_{w,max}$

| $\dfrac{\mathbf{v}_{w,max}}{\delta_{max}}$ | Computational time | Total no. of wear iterations | Max. total wear depth (mm) | Error in wear depth (%) |
|---|---|---|---|---|
| 0.5 % | 19 hr 32 min | 577 | 6.3919 | - |
| 1 % | 16 hr 10 min | 411 | 6.4230 | 0.49 |
| 2 % | 10 hr 17 min | 291 | 6.4849 | 1.45 |
| 5 % | 6 hr 0 min | 183 | 6.6260 | 3.66 |
| 10 % | 3 hr 57 min | 130 | 6.9668 | 8.99 |

## 3.2 Test Case 2: Tuned shrouded bladed disk

A tuned shrouded bladed disk consisting of 40 blades with identical blades is considered, and the fundamental blade sector is as shown in Figure 14 and described in Table 3. This bladed disk is used as a test case to demonstrate the impact of wear on the dynamics with cyclic symmetry property. A fundamental sector consisting of only one blade is assumed with cyclic symmetry boundary conditions to reduce the size of the problem for non-linear analysis. The disk is assumed to be infinitely rigid and so fixed boundary conditions are applied at the blade root. In this way, the cyclic symmetry only operates through the blade-to-blade coupling at the shrouds. Maximum static deflection ($\delta_{max}$) in this case is computed as the maximum deflection at the shroud contact patch for the applied static load in such a way to produce a twisting effect of the blade.

Table 3: Parameters for the shrouded bladed disk and the contact patch

| Parameter | Value |
|---|---|
| Material | Steel |
| Young's modulus ($E$) | 210 GPa |
| Density ($\rho$) | 7860 kg/m$^3$ |
| Blade geometry | As shown in Figure 14 |
| Number of blades | 40 |
| Contact patch dimension | 20mm x 30mm |
| Contact elements | 20 on the left and 20 on the right shroud |
| Tangential contact stiffness ($k_t$) | 93 N/μm |
| Normal contact stiffness ($k_n$) | 113 N/μm |
| Friction coefficient ($\mu$) | 0.5 |
| Wear energy coefficient ($\alpha$) | 2e3 μm$^3$/J |
| Static load ($F_{static}$) | 60 kN |
| Static load/Excitation ratio ($F_{static}/F_{ex}$) | 12 |
| Choice of $\mathbf{v}_{w,max}$ | {0.5%, 1%, 2%, 5%, 10%} of $\delta_{max}$ |

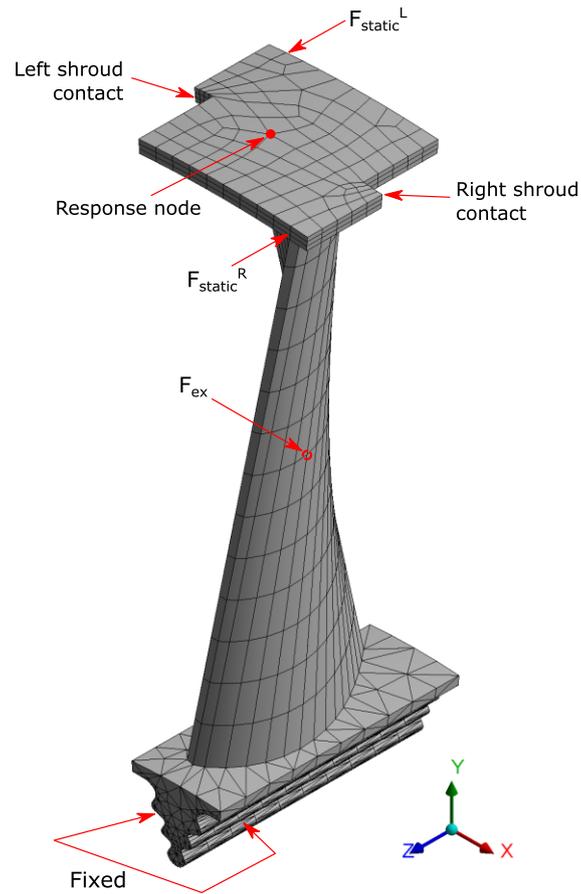

Figure 14: FE model of the blade sector, boundary and loading conditions

Static torque is applied to the blade by two static forces $F_{static}$, to simulate the twisting effect, actually due to the centrifugal force in rotating blades, which produces the static pre-load at the shrouds. A concentrated periodic excitation $F_{ex}$ is also applied at a mid-span node of the airfoil. Figure 15(a) highlights the shroud contact patch consisting of 20 contact elements each side with a 5x4 grid. Figure 15(b) shows the typical contact elements distribution between right contact patch with an adjacent blade left contact patch. The friction coefficient, contact stiffnesses and wear energy coefficient is defined as mentioned in Table 3. The non-linear dynamic analysis is performed by retaining the $0^{th}$ and the $1^{st}$ harmonics in the balance equation. First bending mode with first Engine Order (EO = 1) as shown in Figure 16 is chosen as the frequency of interest to analyse the effect of wear on the dynamics of the bladed disk.

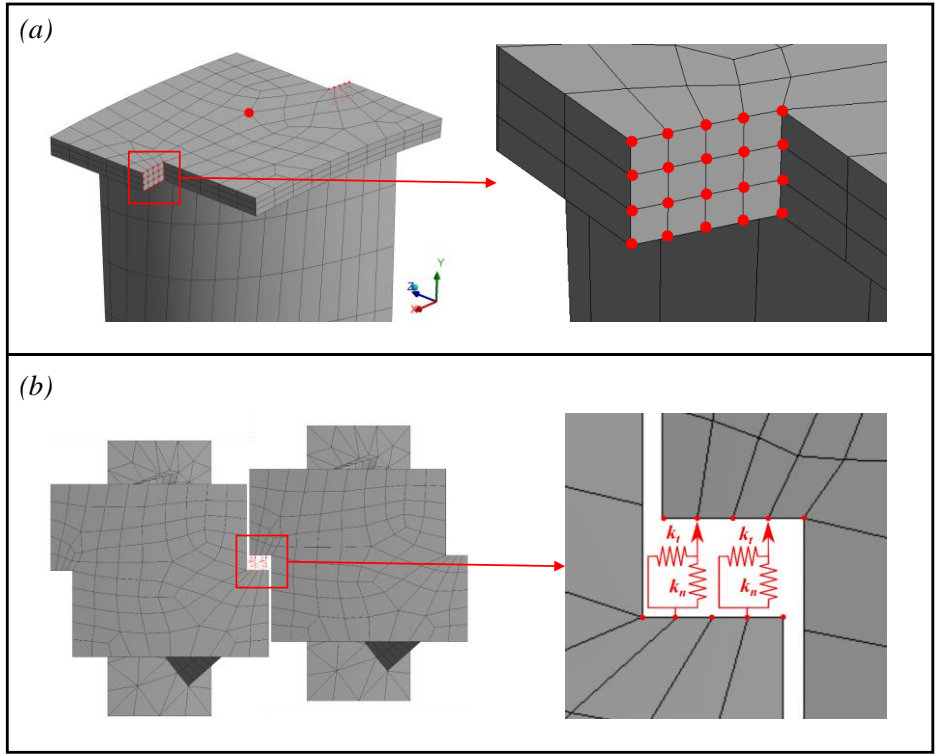

Figure 15: (a) Highlight of the right shroud contact showing 20 contact nodes
(b) Pictorial representation of the applied node-to-node contact model

To define the adaptive wear parameter $\mathbf{v}_{w,max}$, the following logic is applied. The static deflection of the blade is computed under the action of the static forces. Then the average normal displacement of each contact surface is obtained. This value has been set as the maximum possible wear on each surface $\mathbf{v}_{w,max}$, when such a depth is worn out the blades would vibrate freely without any contact with the adjacent blades. Finally, $\mathbf{v}_{w,max}$ is varied from 0.5% to 10% in steps indicated in Table 3 to investigate its effect on the accuracy of the model in predicting the evolution of the system dynamics.

Figure 16 shows the nodal diameter versus frequency relationship up to 15 ND for the stuck case. Figure 17 shows the response for various engine order excitations along with the linear free state and stick state curves for the indicated loads.

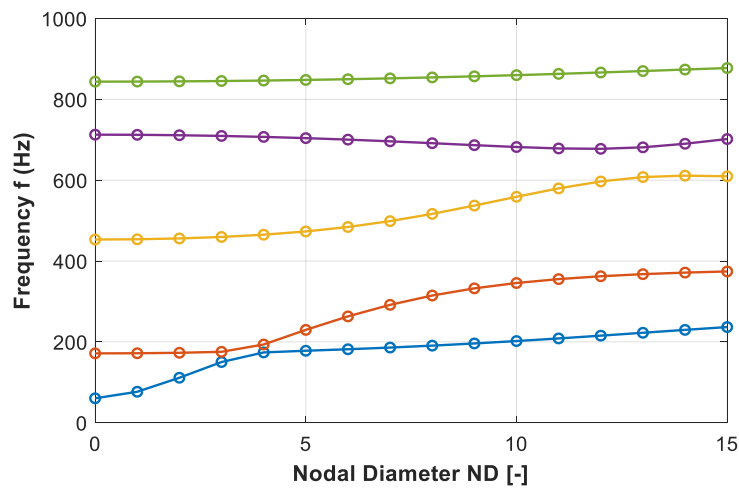

Figure 16: Nodal diameter diagram of the stuck contact shrouded blade

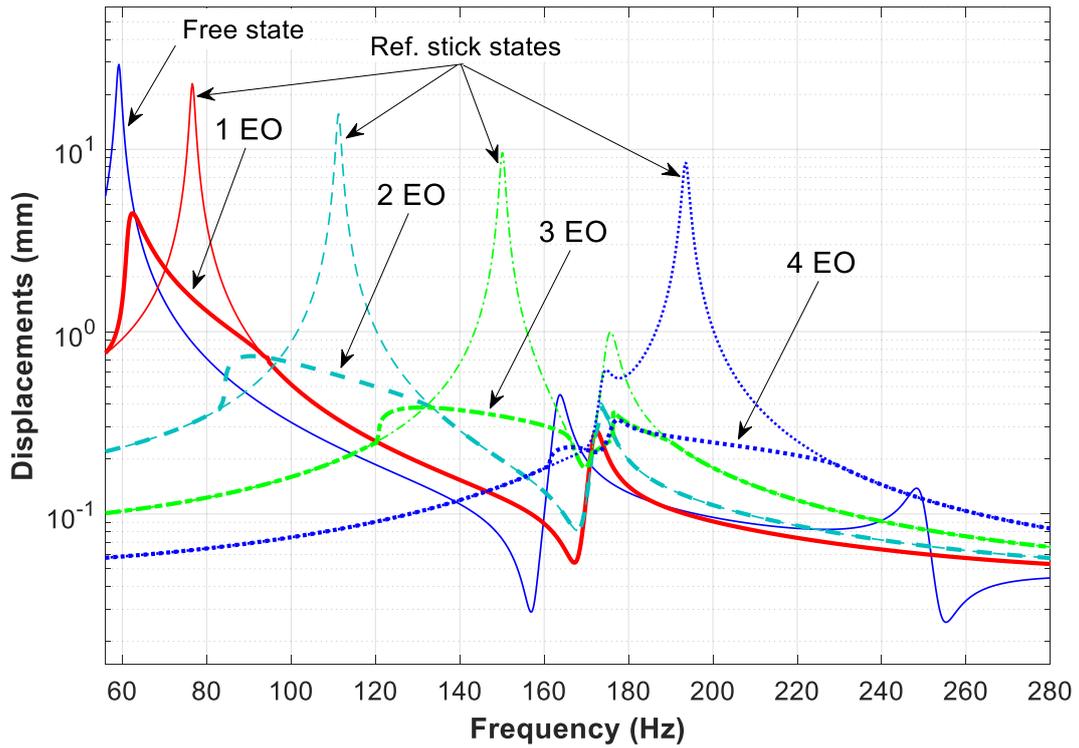

Figure 17: Response to various EO excitations with reference linear free and stick state peaks

Figure 18 shows the non-linear FRF for the given excitation forces with fixed static load and varying excitation loads. The figure also includes a reference linear free state and stick state for comparison. With the increasing excitation force, the resonance frequency moves towards the free state as the contact nodes experience a larger slip.

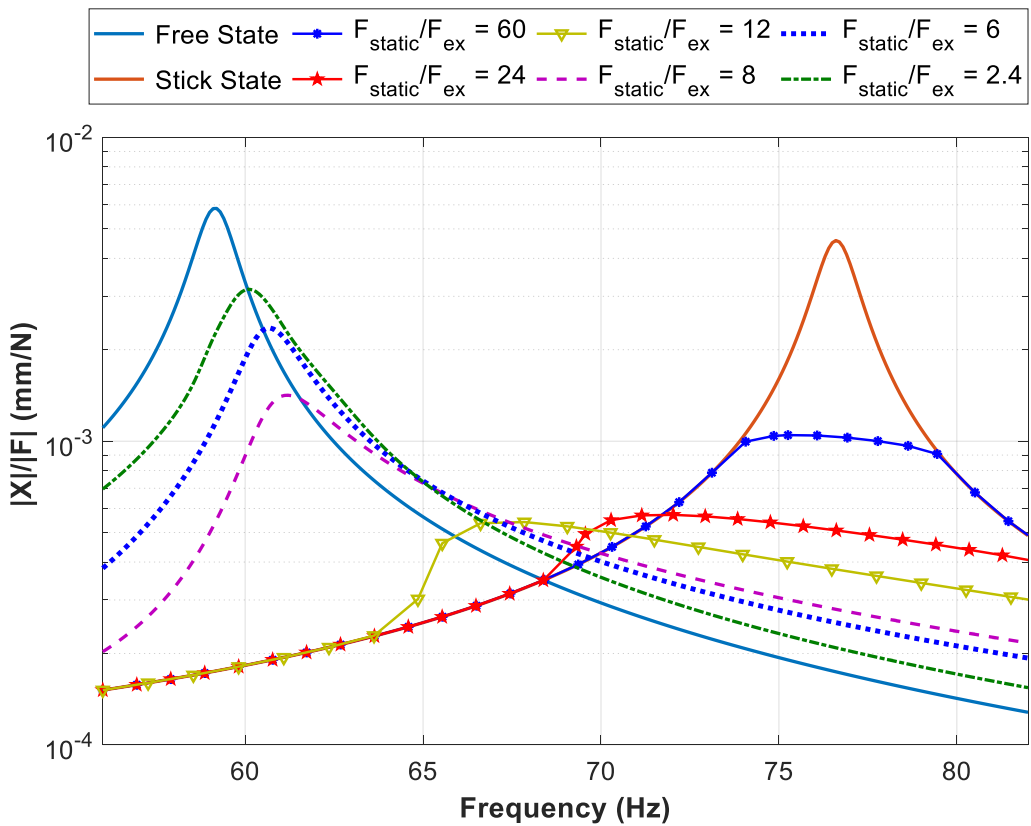

Figure 18: FRF for various excitation loads with a fixed static load (EO = 1)

Figure 19 shows the cumulative wear plot and the contact status plot at various wear iteration intervals. For the given loading conditions, the cumulative wear at the right shroud contact interface, the spatial wear distribution and the contact status at each node can be visualized. At the beginning of the wear, the contact is in the slip-stick state with uniform pressure distribution. As wear progresses, some of the contact nodes are in separation-slip-stick and eventually separation-slip before the complete loss of contact. For simplicity, only one loading condition is shown. The same procedure can be used to visualize various loading conditions and the effect of the adaptive logic parameter $\mathbf{v}_{w,max}$ with an accurate insight at the contact interface behaviour.

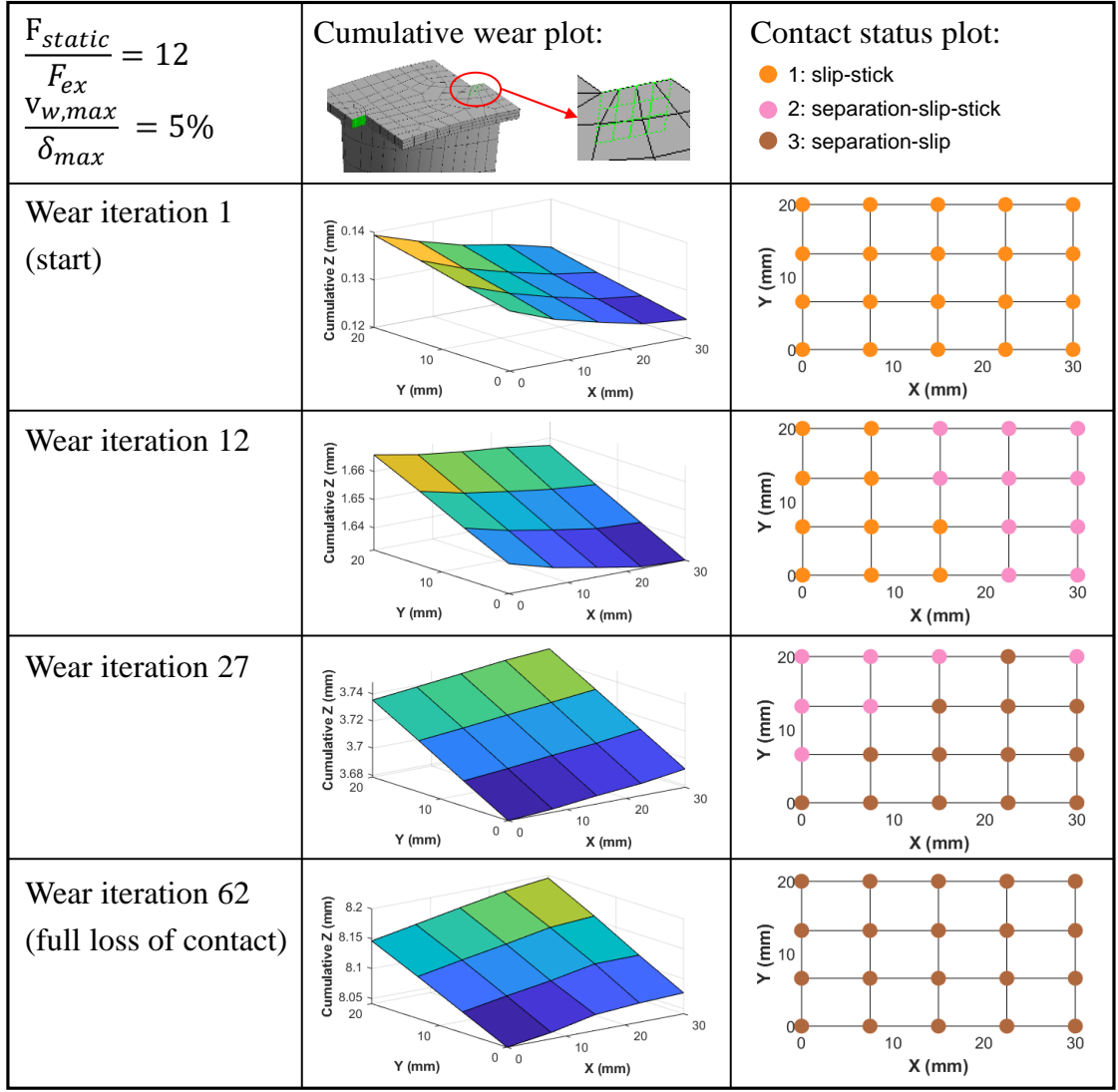

Figure 19: Tabular representation of the cumulative wear plot and contact status plot of the right shroud contact at different wear iterations until the loss of contact for given loading conditions

Figure 20 shows the response with the impact of wear and the backbone of the response curves with the progress of wear. The contact pre-load is continuously changing as the contact is worn and loosens. This loss of contact impacts the dynamic behaviour of the bladed disk. The coupled formulation used in this method is able to effectively update contact pre-load distribution in the non-linear analysis.

Figure 21 shows the number of wear cycles $Z_W$ at each wear iteration as wear progresses until the loss of contact. It is important to note that the number of cycles at each wear iteration is changing. It is adapted according to the geometry and loading conditions. It is worthy of mentioning that the shape of the curve shown in this figure is not unique. Rather it depends on the loading conditions and the pressure distribution at the contact as well as on the relative contact kinematics at the excited resonance.

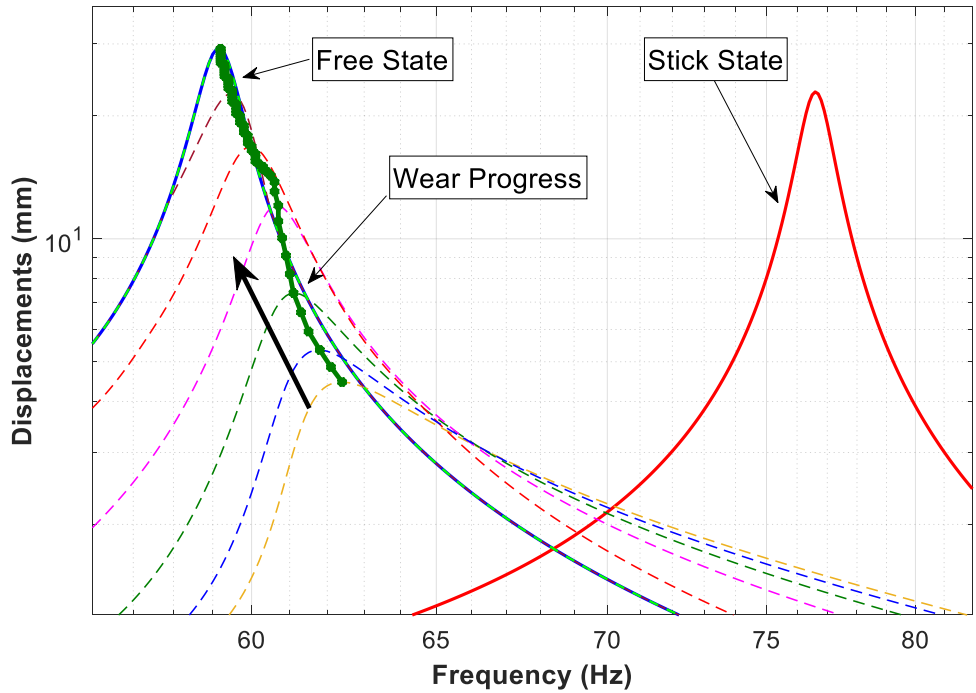

Figure 20: Response graph showing the backbone of the non-linear response (EO = 1) obtained as wear progresses around first bending mode with free and stick states for reference for $v_{w,max}$ = 5% of $\delta_{max}$

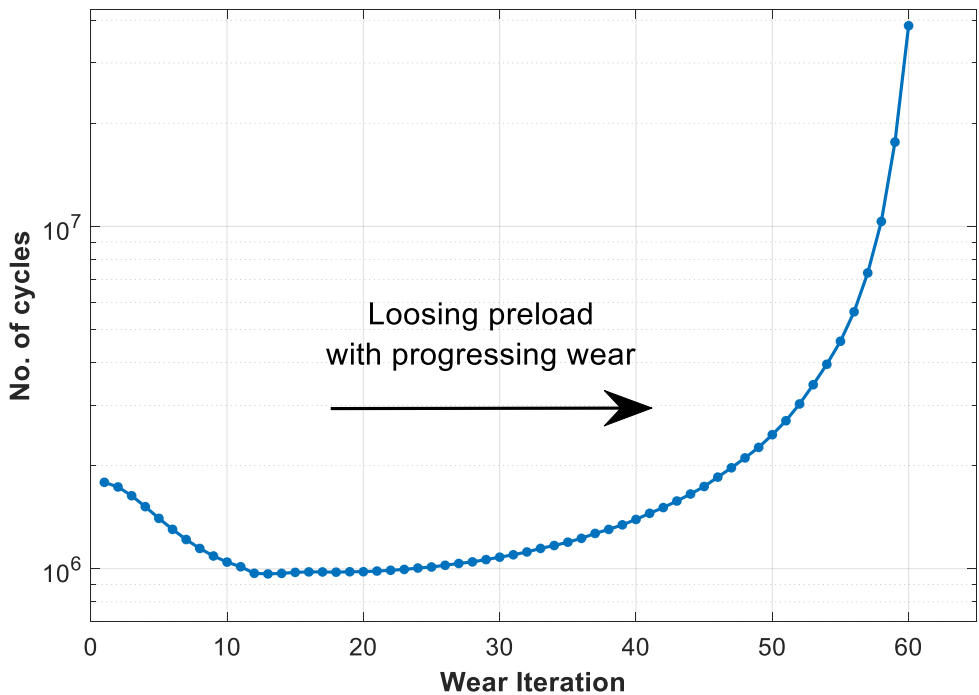

Figure 21: Impact of wear on the number of vibration cycles at each wear iteration until full loss of contact for $v_{w,max}$ = 5% of $\delta_{max}$ [Note: the number of cycles is adaptive at each wear iteration]

It is beneficial to track the maximum response with respect to the cumulative number of cycles as it provides a brief idea of the number of vibration cycles where the amplitude starts to rise considerably. Figure 22 shows the maximum response versus the cumulative number of cycles until the full loss of contact. Results show that for higher values of $v_{w,max}$, the rate of change of the response is under-estimated, leading to non-conservative predictions.

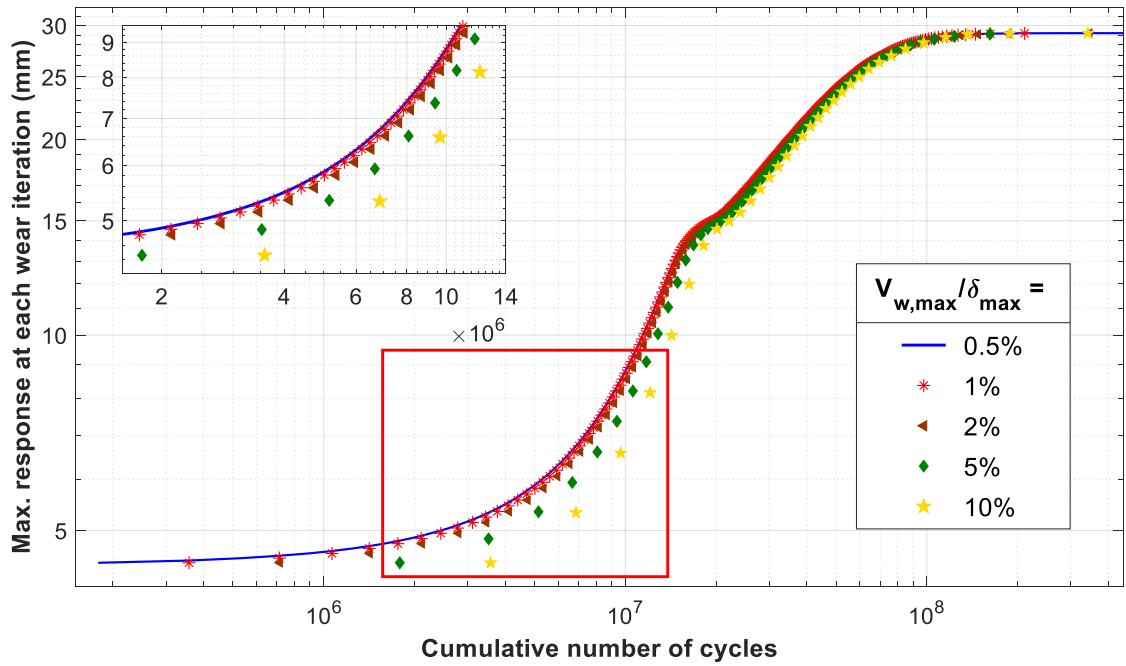

Figure 22: Maximum response at each wear iteration vs the cumulative number of cycles for different values of $\mathbf{v}_{w,max}$

Table 4 summarizes the comparison of the effect of the choice of $\mathbf{v}_{w,max}$ on the computational time, number of wear iterations necessary for the complete loss of contact, maximum wear depth and the error in the maximum wear depth.

Table 4: Comparison of the impact of wear analysis with non-linear dynamic response solver for different values of $\mathbf{v}_{w,max}$

| $\dfrac{\mathbf{v}_{w,max}}{\delta_{max}}$ | Computational time | Total no. of wear iterations | Max. total wear depth (mm) | Error in wear depth (%) |
|---|---|---|---|---|
| 0.5 % | 5 hr 17 min | 596 | 8.2566 | - |
| 1 % | 2 hr 38 min | 298 | 8.2572 | 0.01 |
| 2 % | 1 hr 19 min | 152 | 8.2916 | 0.4 |
| 5 % | 0 hr 35 min | 63 | 8.3808 | 1.5 |
| 10 % | 0 hr 17 min | 34 | 8.4697 | 2.6 |

The indicated computational times in Table 2 and Table 4 is the wall clock time obtained when the non-linear analysis was run on Intel Xeon 4 core processor @ 3.50 GHz and 32 GB RAM standalone workstation.

# 4. Conclusions

The current research work successfully studied the effect of wear on the dynamics of structures with frictional contacts using a multi-scale approach - the wear energy approach coupled with adaptive wear logic for accelerated wear and a coupled static/dynamic harmonic balance method formulation to capture the dynamics. The methodology is demonstrated on two test cases – a cantilever beam with a contact patch and a tuned shrouded bladed disk with shroud contacts with cyclic symmetry properties.

The proposed two-fold aim has been achieved:

1) Modelling the effect of wear on the forced dynamic response and the interface evolution with 'changing contact pre-load'. This was achieved by loading the structure with a static pre-load distribution over the contact and automatically updated during the non-linear analysis as the contact loosens up due to wear. This method eliminates the need for separate non-linear static analysis routine.

2) The effect of the choice of adaptive wear acceleration parameter $\mathbf{v}_{w,max}$ that is described in the paper. The effect of the choice of user-defined parameter $\mathbf{v}_{w,max}$ is studied on the dynamic response of the two test cases. With smaller values of $\mathbf{v}_{w,max}$, the change in frequency and amplitude of response curves, wear evolution, and the contact profile is smoother. But with increased computation time and a higher number of wear iterations. In case of larger values of $\mathbf{v}_{w,max}$, the results are coarse but with faster computation times. The choice of $\mathbf{v}_{w,max}$ is a trade-off between the computation time and the accuracy of results.

## Acknowledgements

This project has received funding from the European Union's Horizon 2020 research and innovation programme under the Marie Sklodowska-Curie grant agreement No 721865.